\documentclass[10pt,twocolumn]{IEEEtran}
\usepackage{booktabs}
\usepackage{url}
\usepackage{graphicx}
\usepackage{float}
\usepackage{cite}
\usepackage{qtree}
\usepackage{color}
\usepackage{tikz}
\usepackage{supertabular}
\usetikzlibrary{arrows,shapes,positioning,shadows,trees}
\tikzstyle{abstract}=[rectangle, draw=black, rounded corners, fill=blue!40, drop shadow, text centered, anchor=north, text=white, text width=2.7cm]
\tikzstyle{comment}=[rectangle, draw=black, rounded corners, fill=green, drop shadow,
    text centered, anchor=north, text=white, text width=3cm]
\tikzstyle{myarrow}=[->, >=open triangle 90, thick]
\tikzstyle{line}=[-, thick]

\begin{document}

\title{QoS in IEEE 802.11-based Wireless Networks:\\ \textit{A Contemporary Survey}}
\author{Aqsa Malik, Junaid Qadir, Basharat Ahmad, Kok-Lim Alvin Yau, Ubaid Ullah.
\thanks{Aqsa Malik, Basharat Ahmad and Ubaid Ullah are students at the School of Electrical Engineering and Computer Science (SEECS) at the National University of Sciences and Technology (NUST), Pakistan. Junaid Qadir is an Assistant Professor at the Electrical Engineering Department of SEECS, NUST. Kok-Lim Alvin Yau is an Associate Professor at the Faculty of Science and Technology, Sunway University, Malaysia}
} 
\maketitle

\begin{abstract}
Apart from mobile cellular networks, IEEE 802.11-based wireless local area networks (WLANs) represent the most widely deployed wireless networking technology. With the migration of critical applications onto data networks, and the emergence of multimedia applications such as digital audio/video and multimedia games, the success of IEEE 802.11 depends critically on its ability to provide quality of service (QoS). A lot of research has focused on equipping IEEE 802.11 WLANs with features to support QoS. In this survey, we provide an overview of these techniques. We discuss the QoS features incorporated by the IEEE 802.11 standard at both physical (PHY) and media access control (MAC) layers, as well as other higher-layer proposals. We also focus on how the new architectural developments of software-defined networking (SDN) and cloud networking can be used to facilitate QoS provisioning in IEEE 802.11-based networks. We conclude this paper by identifying some open research issues for future consideration.

\end{abstract}


\section{Introduction}

The IEEE 802.11 wireless local area networking (WLAN) standard defines one of the most widely deployed wireless technologies in the world. The popularity of wireless networking is driven by the ubiquity of portable mobile hand-held devices, and the convenience of untethered communications. With the increasing deployment of multimedia content on the Internet---such as digital video, voice over IP (VoIP), videoconferencing, and multi-player networked games---along with the deployment of time-sensitive critical applications, there is a strong motivation to develop QoS features to meet the more stringent performance requirements \cite{baghaei2004review}.

While the Internet and data networking models of the IEEE 802.11 WLAN technology, which are based on the \textit{datagram delivery} model of IP, provide simple, adaptive and fault resilient network, they are ill-suited to QoS provisioning. The underlying datagram model of IP is a \emph{best-effort} service---i.e., while the network tries to deliver packet to the destination correctly without any packet losses, it makes no guarantees. Multimedia applications, in particular, need stronger guarantees about the minimum throughput and maximum latency to work satisfactorily. An expensive solution for ensuring QoS is to \textit{overprovision}. Most of the Internet QoS effort has focused on how to get a network with less capacity meet application requirements at a lower cost. In IEEE 802.11 WLANs, the error and interference prone nature of wireless medium---due to fading and multipath effects \cite{gast2005802}---makes QoS provisioning even more challenging. The combination of best-effort routing, datagram routing, and an unreliable wireless medium, makes the task of QoS provisioning in IEEE 802.11 WLANs very challenging.

In this survey, we provide a focused overview of work done to ensure QoS in the IEEE 802.11 standard. We have the following three goals: (i) to provide a self-contained introduction to the QoS features embedded in the IEEE 802.11 standard; (ii) to provide a layer-wise description and survey of techniques adopted for ensuring QoS in the IEEE 802.11 networks; and (iii) to survey the applications of new networking architectures---such as software defined networks (SDN) and cloud computing---for QoS provisioning in the IEEE 802.11-based WLANs.

\vspace{2mm}
\textit{Contributions of this paper:} A lot of research has been conducted on the topic of QoS \cite{wang2001internet}, including numerous surveys that have focused on the QoS problems for specific wireless networks such as wireless sensor networks (WSNs) \cite{chen2004qos}, wireless mesh networks (WMNs) \cite{mogre2007qos}, and IEEE 802.11-based WLANs \cite{zhu2004survey} \cite{lindgren2001evaluation} \cite{ni2004survey} \cite{aboul2009wireless}. Our work is different from the previous work in that we provide an updated account of QoS literature in IEEE 802.11-based wireless networks including a discussion of recent architectural developments, such as cloud computing and SDN, that facilitate finer network management control, so we have reinvigorated the interest of the research community in providing high QoS in IEEE 802.11-based wireless networks. In addition to highlighting the QoS features incorporated into the IEEE 802.11 networking standard, we also highlight different QoS approaches pertaining to different layers of the TCP/IP model. 


\vspace{2mm}	
\textit{Organization of this paper:} This survey is organized in the following way. In Section \ref{sec:InternetQoS}, we provide a broad-based introduction on the general area of Internet QoS. In Section \ref{sec:techAdvancement}, we present the QoS features which are intrinsic to IEEE 802.11\footnote{We note here that the IEEE 802.11 standard directly addresses the PHY and MAC layers only.} for the physical (PHY) and the medium access control (MAC) layers. We provide an overview of the IEEE 802.11 a/b/g and higher-throughput IEEE 802.11 standards (802.11 n/ac/ad) in Sections \ref{sec:802.11abg} and \ref{sec:802.11nadac}. We follow this by a discussion on MAC layer QoS features proposed in IEEE 802.11 in Section \ref{sec:MAC}. Apart from the QoS features that are part of the IEEE 802.11 standard, various work has focused on QoS improvement including work at the network layer (discussed in Section \ref{sec:NET}), the transport layer (discussed in Section \ref{sec:Transport}), the application layer (discussed in Section \ref{sec:Application}) as well as cross-layer work (discussed in Section \ref{sec:CrossLayer}). The promise of recent architectural developments, such as cloud computing and SDN, in enabling QoS, along with a survey of proposed work, is provided in Section \ref{sec:modernArch}. Thereafter, we discuss some open research issues in Section \ref{sec:futurework}. Finally, we provide concluding remarks in Section \ref{sec:conclusion}. 

\vspace{5mm}
To facilitate the reader, acronyms used in this paper are collected in Table \ref{tab:acronyms} as a convenient reference.

\begin{table}
\caption{Acronyms used in this paper.}
\label{tab:acronyms}
\footnotesize
\begin{tabular}{p{1.5cm} p{6cm}}
\toprule
\textbf{\textit{Acronym}} & \textbf{\textit{Expanded Form}}\\
\midrule

AMC & Adaptive Modulating Scheme\\
APSD & Automatic Power Save Delivery\\
ARQ & Automatic Repeat reQuest\\
ATM & Asynchronous Transfer Mode\\
BER & Bit Error Rate\\
BPSK & Binary Phase Shift Keying\\
CAC & Call Admission Control\\
CAP & Control Access Period\\
CbWN & Cloud-based Wireless Network\\
CW & Contention Window\\
DCF & Distributed Coordination Function\\
DFS & Distributed Fair Scheduling\\
DIFS & DCF Interframe Space\\
DiffServ & Differentiated Services\\
DSSS & Direct-Sequence Spread Spectrum\\
EDCA & Enhanced Distributed Channel Access\\
EDCF & Extended DCF\\
EDD & Earliest Due Date\\
FEC & Forward Error Correction\\
FHSS & Frequency Hopping Spread Spectrum\\
FCFS & First-Come First-Served\\
FIFO & First In First Out\\
HCF & Hybrid Coordination Function\\
HCCA & HCF Controlled Channel Access\\
IntServ & Integrated Services\\
LTE & Long-Term Evolution\\
MAC & Media Access Control\\
MDP & Markov Decision Process\\
MPDU & MAC Protocol Data Unit\\
MPLS & Multiprotocol Label Switching\\
MIMO & Multiple Input Multiple Output\\
NUC & Network Utilization Characteristic\\
OMAR & Opportunistic Medium Access and Adaptive Rates\\
OSAR & Opportunistic Scheduling and Auto Rate\\
PCF & Point Coordination Function\\
PIFS & PCF Interframe Spacing \\
PHB & Per-Hop Behaviour\\
PSTN & Public Switched Telephone Network\\
QAM & Quadrature Amplitude Modulation\\
QoE & Quality of Experience\\
QoS & Quality of Service \\
RTS & Request To Send\\
SDN & Software Defined Networking\\
SIFS & Short Interframe Spacing\\
SISO & Single Input Single Output\\
STA & (Wireless) Station \\
SWN & Software Defined Wireless Network\\
TDM & Time Division Multiplexing\\
TXOP & Transmission Opportunity\\
VoIP & Voice over IP\\
VM & Virtual Memory\\
WFQ & Weighted Fair Queuing\\
WLAN & Wireless Local Area Networking\\
WRR & Weighted Round Robin\\
WSN & Wireless Sensor Network\\
\bottomrule
\end{tabular}
\end{table}

\section{Internet QoS---A Broad Introduction}
\label{sec:InternetQoS}

There has been a lot of work on Internet QoS, the bulk of which has focused on wired networks \cite{Wang2001} \cite{jha2002engineering}. While many of the ideas developed for Internet QoS are also applicable more broadly to wireless QoS, wireless networks do provide some unique challenges motivating the development of new methods \cite{setton2005cross}. In this section, we provide a broad overview of the abundant literature on Internet QoS.


The original applications of the Internet---such as file transfer and email---are \emph{elastic} applications which are not bound by stringent performance requirements, and therefore match well with the Internet's datagram delivery model. The modern Internet world, which is full of multimedia applications, requires QoS guarantees that users have come to expect from the telecommunications networking world. To support multimedia and other interactive/high performance applications, there is a need to support QoS features through \textit{QoS provisioning} that provides resource assurance along with service differentiation. Various techniques have been developed to facilitate QoS provisioning, including (i) congestion control, (ii) admission control, and (iii) traffic shaping and engineering.

In the remainder of this section, we study the problems of resource allocation and service differentiation, and will introduce the techniques of admission control, congestion control, scheduling, as well as traffic shaping and engineering that can be used to facilitate QoS provisioning.

\subsection{Resource Allocation}

Fundamentally, many QoS issues stem from the problem of \textit{resource allocation}. A computer network is composed of various resources---such as links of varying bandwidths, routers with varying buffer sizes---that are shared by the different network applications and users. Packet delays and losses occur if the network resources cannot meet all the traffic demands. A network that supports QoS must actively manage resource allocation to satisfy various users' and applications' demands. Without appropriate resource allocation, network performance and service quality deteriorate rapidly under heavy load due to dropped packets and congestion. There are two main architectural approaches to resource allocation in the Internet: Integrated Services (IntServ) and Differentiated Services (DiffServ). Apart from IntServ and DiffServ, other QoS frameworks have also been proposed \cite{aurrecoechea1998survey}. We, however, focus only on the more important QoS framework proposals, namely IntServ and DiffServ.

\vspace{2mm}
\subsubsection{IntServ}

IntServ performs \emph{per-flow resource reservation} for service differentiation. IntServ provides services on a per-flow basis where a flow is a packet stream with common source address, destination address and port number. In IntServ, a packet scheduler is used to enforce resource allocation to individual flows while supporting prioritization. The IntServ scheduler can be used to provide delay bounds. The delay bounds can be deterministic or statistical---for deterministic bounds, isolation or dedication of resources is required, while statistical bounds can be provided when statistical multiplexing \cite{mahadevan1999quality} is used. There are two key IntServ abstractions, namely Reserved Resources and Standard Resources. In the Reserved Resource abstraction, the router must know the amount of resources currently reserved for on-going sessions. The Standard Resource abstraction includes the capacities of the links and the router buffers, respectively. An example is Call Setup in which buffers are kept at the routers. These buffers ensure a specific amount of bandwidth is allocated to the flows at each router \cite{{zhang1993rsvp}} \cite{{clark1992supporting}} \cite{{shenkerspecification}} \cite{{braden1994integrated}}.

\vspace{2mm}
\subsubsection{DiffServ}

DiffServ, on the other hand, performs \emph{per-class resource reservation} for service differentiation, and makes use of prioritization, multiple forwarding classes, and edge policing to categorize traffic into different classes; and the traffic is treated according to its respective classes. The edge routers are responsible for the complex operations in the network; while the core routers perform simple and easy computations. The packet-handling rule in DiffServ is termed as Per-Hop Behaviour (PHB). In other words, each network device along a path behaves in a certain way in which a specific group of packets have the same priority value. The PHB rule decides whether a packet needs to be forwarded or dropped depending on the QoS-based precedence value of the packet. However, the framework is very complex and cannot be applied to heterogeneous networks \cite{blake1998architecture} \cite{jacobson1999expedited} \cite{nichols1999two}. DiffServ has been used for implementing QoS in various IEEE 802.11-based wireless networks such as \cite{chaouchi2004adaptive} \cite{garcia2003quality}.

\subsection{Service Differentiation}

Service differentiation is used to support multiple services with diverse requirements---such as interactive delay-sensitive services along with elastic delay-tolerant file transfer services \cite{christin2003qos}. The overprovisioning of network resources is not always possible in radio networks, thus making service differentiation an integral component of most QoS-based solutions. In service differentiation, several parameters (e.g., packet deadline) can be modified to define how a flow should access the wireless medium \cite{chen2004qos}. A variety of services can be provided by the use of simple network parameters deployed in network nodes, and these services can be classified according to a large number of characteristics \cite{aad2001differentiation}. The QoS of the system is enhanced by differentiating the priority of each host and offering them different levels of QoS parameters.

\begin{figure}[t]
\begin{center}
\includegraphics[width=.5\textwidth]{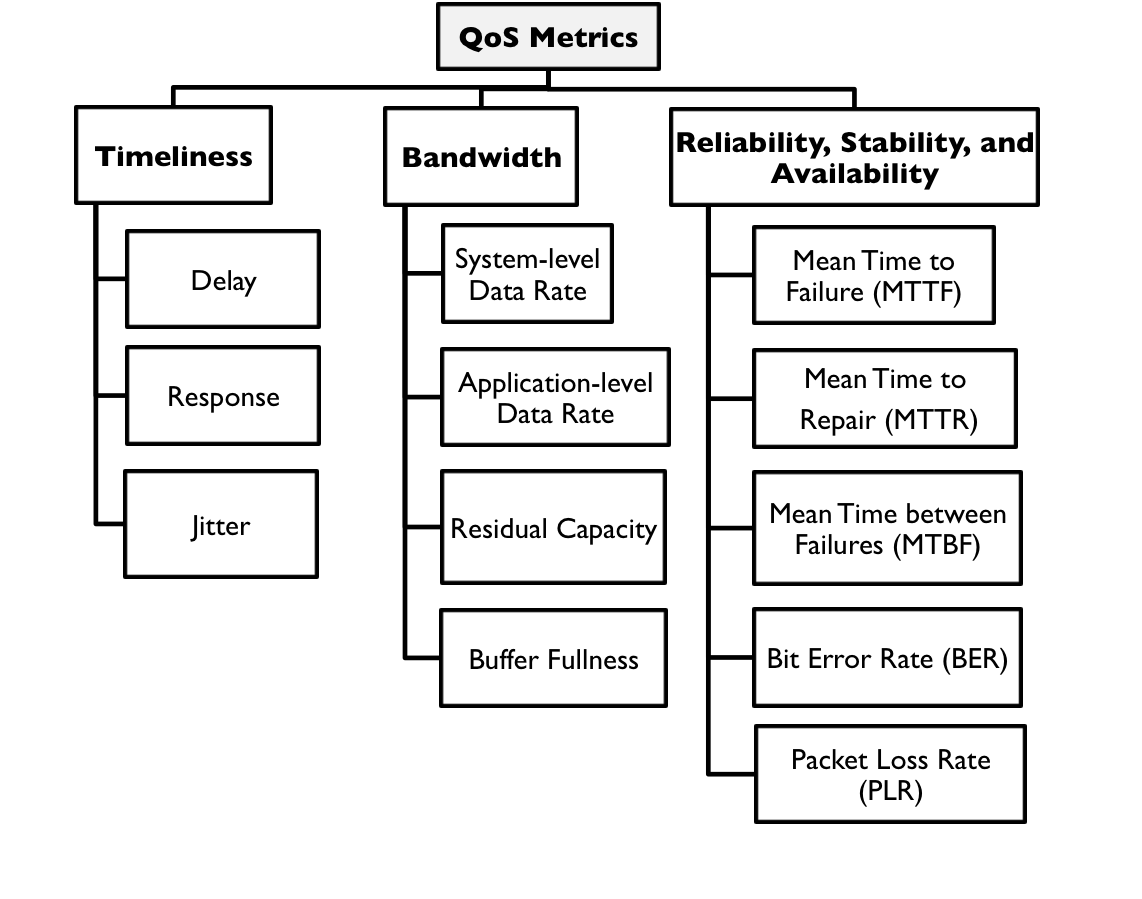}
\caption{Taxonomy of QoS metrics (adapted from \cite{hanzo2007survey} and \cite{chalmers1999survey}).}
\label{fig:metrics}
\end{center}
\end{figure}

Service requirements are often application-specific. For example, certain applications are delay-sensitive (e.g., voice conferencing which is sensitive to round-trip delay), while others are concerned more with average transmission rate (e.g., bulk file transfer). Service requirements are often expressed using metrics (i) bandwidth, (ii) delay, (iii) jitter, and (iv) loss rate. A more comprehensive, but still non-exhaustive, listing of QoS metrics is displayed in Figure \ref{fig:metrics}. To accommodate the impact of these metrics, the network must support multiple QoS strategies to support different applications \cite{ksentini2004adaptive} \cite{jha2002engineering}.

The bandwidth requirements of different applications are different. Some applications, such as email, remote login and audio, require less bandwidth, while video and file transfers require high bandwidth. Similarly, the delay requirements also differ with the type of application. Some applications, such as email, are not delay-sensitive. However, interactive applications, such as web browsing, videoconferencing and live streaming, have more stringent delay requirements. The variation in the packet arrival time within a stream of packets is called jitter. Email, file sharing and remote login are not affected by jitters in the network traffic; while real-time applications, audio and video do. The loss of packets in the case of audio and video is not of significance. The other applications, as discussed previously, cannot tolerate an increased packet loss rate in transmissions \cite{tanenbaum2003computer}. In this work, we will focus mostly on the QoS parameters of delay, throughput, and guaranteed bandwidth.

\subsection{Admission Control}

One way of supporting QoS is through \textit{admission control}---in which new sessions are allowed onto the network only if sufficient resources are available to provide service to the new and existing sessions \cite{knightly1999admission}. The interest in the field of admission control, has been driven by the idea that regulation of incoming traffic flows prevents network congestion, and helps in ensuring QoS. Call Admission Control (CAC) is a traffic management system employed in public switched telephone networks (PSTNs) \cite{perros1996call}. The CAC scheme is easy to implement because of the homogeneous environment. However, the present networking environment of homogeneous network is not preserved. Hence, the admission control function is more challenging in heterogeneous networks (e.g., joint WLANs and IP networks) \cite{gao2005admission}. A new flow request is admitted only if the ongoing flows are not negatively affected. The common parameters used for admission control are peak bandwidth requirement and the average rate \cite{aboul2009wireless}. Therefore, the task of admission control is to maximize resource utilization in the network, and to control the amount of traffic to achieve the predefined performance objectives of the current flows. 

Hou et al. \cite{hou2009theory} have presented a formal theory of QoS provisioning in unreliable wireless networks, such as the IEEE 802.11-based wireless networks, which subsumes a framework for jointly addressing three important QoS criteria, namely delay, delivery ratio, and channel reliability. They also propose algorithms and policies for admission control and scheduling that can be implemented in IEEE 802.11-based networks. The authors analytically develop necessary and sufficient conditions to satisfy these three criteria. More details of admission control techniques in the context of IEEE 802.11 standard are described later in Section \ref{sec:NET}.

\subsection{Congestion Control}

Congestion control in the modern Internet is typically performed using the TCP protocol \cite{vicisano1998tcp}. Congestion in a network may occur if the number of packets sent to the network is greater than the number of packets a network can handle. Congestion control refers to the techniques to control the congestion level and keep the load below the capacity. In the QoS-integrated services, the congestion control mechanism should be different for different kinds of sources: e.g., file transfer/ email is different from real-time voice/video applications \cite{xiao1999internet}. The QoS enabled routers provide services to certain flows based on their requirements. Congestion control helps to provide priority differentiation of flows by servicing queues in different manners (e.g., the order in which the flows are serviced).

\subsection{Scheduling}
Scheduling is the key to share network resources fairly among users in a network, and it provides service guarantees to time-critical applications. The scheduler first decides the order of requests to be served, and then it manages the queues of these awaiting requests. The scheduling scheme is important for the networks because there are two types of applications. One is insensitive to the performance that users receive from the network, and the other has a strict bound on the performance. The scheduling can provide different services to the flows using parameters such as different \emph{bandwidths}---by serving only a single flow at a particular interval; different \emph{mean delays}---according to the level of priority defined for the flow; and different \emph{loss rates}---by assigning more or fewer buffers to the flows \cite{keshav1997engineering}. The scheduling mechanism adopted in the IEEE 802.11 standard is explained in detail later in Section \ref{sec:MAC}.

\subsection{Traffic Shaping and Engineering}
Traffic in data networks is bursty in nature. Traffic shaping is a technique for handling the bursty nature of the traffic entering a network through controlling and allocating appropriate levels of network bandwidth \cite{tanenbaum2003computer}. The goal is to regulate average traffic rate and reduce congestion. The traffic shaping is performed at the boundary nodes. These nodes have classifiers that mark the flows according to their service requirements. The mechanisms of traffic management can be classified in a number of ways \cite{wang2001internet}. One possible criterion is time scale \cite{aboul2009wireless}. In order to achieve QoS guarantees, decisions on buffering and forwarding must be performed quickly. Traffic engineering is the process that maximizes network utilization through careful distribution of network resources \cite{wang2001internet}. Most of the Internet backbones currently rely on label switching by adopting `multi protocol label switching' (MPLS) technology. The purpose of label switching is to enhance the scope of traffic engineering, QoS provisioning and overlay networks \cite{peterson2007computer}. The traffic shaping mechanism for the IEEE 802.11 standard is defined in Section \ref{sec:MAC}.

\section{QoS Support in IEEE 802.11}
\label{sec:techAdvancement}

Standards in the IEEE 802 project target the PHY layer and the MAC layer. While IEEE 802.3 defines the PHY and MAC layers for wired LANs; the prominent IEEE 802.11 standard, which is the focus. The first IEEE 802.11 specification was published in 1997, and it has undergone numerous subsequent amendments. The IEEE 802.11 working group has various task groups focusing on a myriad of niche concerns with an elaborated description of the IEEE 802.11 universe provided in \cite{hiertz2010ieee}. The focus of the various task groups is summarized in tabulated form in Tables \ref{tab:802.11 family} and \ref{tab:802.11 family active}. 

\begin{table*}
\scriptsize
\caption{The IEEE 802.11 Standard Task Groups With Completed Specifications.}
\renewcommand{\arraystretch}{1.35}
\centering
\begin{tabular}{|p{1.0cm}|p{4cm}|p{2.5cm}|p{7.7cm}|}
\hline
\textbf{\textit{Task Group}} & \textbf{\textit{Title}} & \textbf{\textit{Status}}& \textbf{\textit{Comment}}\\
\hline
802.11a & Higher Speed PHY Extension in the 5 GHz Band & Completed; published as IEEE Std. 802.11a-1999 & Defines a PHY to operate in the UNII band. \\
\hline
802.11b & Higher Speed PHY Extension in the 2.4 GHz Band & Completed; published as IEEE Std. 802.11b-1999 & Supports a higher rate PHY in the 2.4 GHz band. \\
\hline
802.11d & Operation in Additional Regulatory Domains & Completed; published as IEEE Std. 802.11-2007 & Allows devices to comply with regional requirements.\\
\hline
802.11e & MAC layer enhancements for QoS & Completed; published as IEEE Std. 802.11-2007 & Enhances the IEEE 802.11 MAC to improve and manage QoS.\\
\hline
 802.11g & Further Higher Data Rate Extension in the 2.4 GHz Band & Completed; published as IEEE Std. 802.11-2007 & Provides higher speed PHY extensions to the IEEE 802.11b standard.\\
\hline
802.11h & Spectrum and Transmit Power Management Extensions in the 5 GHz Band & Completed; published as IEEE Std. 802.11-2007 & Defines \emph{dynamic frequency selection} (DFS) and \emph{transmitter power control} (TPC) for the purposes of efficient spectrum sharing and energy consumption. \\ 
\hline
802.11i & MAC Security Enhancements & Completed; published as IEEE Std. 802.11-2007 & Enhances IEEE 802.11 MAC to provide security, privacy and authentication mechanisms by improving the wired equivalent privacy (WEP) protocol.\\
\hline
802.11j & 4.9 GHz Operation in Japan & Completed; published as IEEE Std. 802.11-2007 & Operates in the 4.9 to 5 GHz band to conform to the Japanese radio regulations.\\
\hline
 802.11k & Radio Resource Management & Completed; published as IEEE Std. 802.11-2012 & Provides interfaces to higher layers for radio resource management and network measurements.\\
\hline
802.11n & Enhancements for Higher Throughput & Completed; published as IEEE Std. 802.11-2012 & Provides improvements to the IEEE 802.11 standard to provide high throughput (greater than 100 Mbps).\\
\hline
802.11p & Wireless Access in Vehicular Environments (WAVE) & Completed; published as IEEE Std. 802.11-2012 & Provides car-to-car communication, with the aim to enhance the mobility and safety of all forms of surface transportation, including rail and marine.\\
\hline
802.11r & Fast Roaming/Fast BSS Transition & Completed; published as IEEE Std. 802.11-2012 & Provides continuous connectivity, as well as fast and seamless hand-off across wireless devices in motion. \\ 
\hline
802.11s & WLAN Mesh Networks & Completed; published as IEEE Std. 802.11-2012 & Enhances the IEEE 802.11 standard to support wireless mesh networking (WMN). \\
\hline
802.11u & Interworking with External Networks & Completed; published as IEEE Std. 802.11-2012 & Provides convergence to IEEE 802.11 and GSM by allowing multi-mode phones to join an IEEE 802.11 WLAN.\\
\hline
802.11v & Wireless Network Management & Completed; published as IEEE Std. 802.11-2012 & Extends the IEEE 802.11 PHY and MAC layers to provide network management for STAs.\\
\hline
802.11w & Protected Management Frames & Completed; published as IEEE Std. 802.11-2012 & Defines security mechanisms for management frames.\\
\hline
802.11y & Contention-based Protocol & Completed; published as IEEE Std 802.11y-2008 & Provides contention-based protocols for operation in the 3.65 GHz band in the USA.\\
\hline
802.11z & Extensions to Direct Link Setup & Completed; published as IEEE Std 802.11z-2010 & Provides an AP-independent direct link setup.\\
\hline
802.11aa & Video Transport Stream & Completed; published as IEEE Std 802.11z-2010 & Defines various MAC enhancements for robust audio video streaming.\\
\hline
802.11ac & Very High Throughput WLAN & Completed; published as IEEE Std 802.11ac-2013 & Provides high throughput (greater than 1 Gbps) operation in bands below 6 GHz.\\
\hline
802.11ad & Very High Throughput WLAN operating in 60 GHz & Completed; published as IEEE Std 802.11ad-2012 & Provides high throughput (greater than 1 Gbps) operation in 60 GHz band.\\
\hline
802.11ae & Prioritization of Management Frames & Completed; published as IEEE Std 802.11ae-2012 & Defines mechanisms for prioritizing IEEE 802.11 management frames using existing mechanisms for medium access.\\
\hline
802.11af & Wireless LAN in the TV White Space & Completed; published as IEEE Std 802.11af-2013 & Defines legal requirements for channel access and coexistence in the TV white space.\\
\hline

\end{tabular}
\label{tab:802.11 family}
\end{table*}

\begin{table*}
\scriptsize
\renewcommand{\arraystretch}{2}
\caption{The IEEE 802.11 standard \emph{active} task groups.}

\centering
\begin{tabular}{|p{1cm}|p{3.5cm}|p{2.3cm}|p{8.5cm}|}
\hline
\textbf{\textit{Task Group}} & \textbf{\textit{Title}} & \textbf{\textit{Status}}& \textbf{\textit{Comment}}\\
\hline
802.11m & IEEE 802.11 Standard Maintenance and Revision & Active; published as IEEE Std 802.11-2012 & Provides maintenance for the IEEE 802.11 standard by rolling published amendments into revisions of the IEEE 802.11 standard.\\
\hline
802.11ah & Operation in Sub 1 GHz Frequencies & Active & Supports applications that benefit from range extension, such as smart meters.\\
\hline
802.11ai & Fast Initial Link Set-up & Active & Reduces time for a WLAN client to securely setup an association in less than 100ms.\\
\hline
802.11aj & Very High Throughput & Active & Operates in the millimeter-wave bands in China.\\
\hline
802.11ak & Enhancements for Transit Links Within Bridged Networks & Approved \newline(December 2012) & Provides protocols and procedures to enhance the ability of IEEE 802.11 media through bridging by using IEEE 802.1 mechanisms across an IEEE 802.11 link.\\
\hline
802.11aq & Pre-association Discovery (PAD) & Approved \newline(December 2012) & Defines modifications to the IEEE 802.11 standard, including layers above the PHY layer, in order to enable delivery of pre-association service discovery information by IEEE 802.11 stations.\\
\hline
802.11ax & High-efficiency Wireless LAN & Approved \newline(March 2014) & Improving spectrum efficiency, area throughput and real world performance in indoor and outdoor deployments.\\
\hline

\end{tabular}
\label{tab:802.11 family active}
\end{table*}

Providing QoS services in applications using traditional IEEE 802.11 standards is difficult since they provide no explicit mechanisms for service differentiation. Various parameters extracted from the general traffic layout are used to ensure acceptable QoS in these networks, such as \textit{goodput}---which is the measure of packet arrival rate during a fixed period of time; \textit{load level}---which indicates the usage of a medium on per time basis; and \textit{available bandwidth}---which measures the rate at which new flows can send traffic without affecting the existing flows in the network \cite{dujovne2010taxonomy} \cite{ng2005experimental}. The legacy standards only have the basic distributed coordination function (DCF) and the optional point coordination function (PCF) enhancements, such as collision avoidance and a first in first out (FIFO) scheduler \cite{ni2005performance}. Even with the contention-free PCF, the QoS problem could not be solved. Consequently, the services provided to the users do not have optimal performances for various applications including audio and video applications during heavy network loads \cite{wu2001streaming}. In Section \ref{sec:802.11e}, we present the IEEE 802.11e standard, which provides MAC layer enhancements for QoS, that incorporates traffic priority and queueing to enable service differentiation among the flows \cite{Mangold2002}.

\subsection{QoS Support in IEEE 802.11 a/b/g/n}
\label{sec:802.11abg}

The IEEE 802.11 has evolved in different eras to satisfy differing requirements of applications \cite{lamaire1996wireless} \cite{raniwala2005architecture}. The IEEE 802.11a standard is part of the original IEEE 802.11 standard, and it operates in the 5 GHz range with a data rate of 54 Mbps while supporting the frequency hopping spread spectrum (FHSS) and direct sequence spread spectrum (DSSS). Unfortunately, due to the use of high frequency spectrum, the technique of DSSS faced the problems of short transmission range and interference \cite{zhou2006crowded}. To address this problem, IEEE 802.11b, which is also based on DSSS technology, operates in the 2.4 GHz spectrum with a data rate of 11 Mbps. The 802.11b standard is not backward compatible with the IEEE 802.11a standard. The IEEE 802.11g standard---operating at 2.4 GHz with a data rate upto 54 Mbps---is introduced for backward compatibility with the IEEE 802.11a standard. 

\emph{QoS limitations of IEEE 802.11a/b/g/n}: In the DCF-based schemes, the access to the medium is given on first come first served (FCFS) basis. This creates some fairness problems, which can potentially result in flows being deprived of their fair bandwidth share. There is no proper mechanism to distinguish between the flows on priority basis in the PCF-based environments. The legacy standards of IEEE 802.11 a/b/g/n have no standard mechanisms to ensure QoS \cite{mohapatra2003qos} \cite{guo2002class}. Because these standards do not incorporate admission control, performance degradation occurs during heavy traffic load.

\subsection{QoS Support in High Throughput IEEE 802.11 (802.11 n/ac/ad)}
\label{sec:802.11nadac}


To support the need of high throughput wireless networking, various IEEE 802.11 standards have been proposed in recent time such as the 802.11n, 802.11ac, and 802.11ad standards. The IEEE 802.11n standard is based on the multiple input multiple output (MIMO) technology, and it offers a high data rate of upto 600 Mbps. The IEEE 802.11ac standard aims to support an even higher data rate of 1 Gbps, while IEEE 802.11ad standard aims at achieving a rate upto 7 Gbps exploiting the wideband channels available in the 60 GHz band. These new standards incorporate scheduling mechanisms, call admission control algorithms, and PHY and MAC layer enhancements for supporting multimedia applications with QoS. The interested reader is referred to a comprehensive study of QoS support in very high throughput IEEE 802.11 architectures (IEEE 802.11 n/ac/ad) presented in \cite{charfi2013phy}.

\subsection{MAC layer QoS Features for IEEE 802.11}
\label{sec:MAC}

In general, the major techniques used for ensuring QoS at the MAC layer include admission control and scheduling. In the IEEE 802.11 standard, the MAC layer provides the functionality of addressing, framing, reliability check, and access coordination to the wireless medium \cite{li2007real}. The MAC layer with QoS enhancements aims to provide the network with a much reduced overhead, segregating frames on the priority basis, and keeping the collisions to the least possible level. This section describes the techniques implemented in the legacy IEEE 802.11 standard. The rest of this subsection presents a description of the QoS-focused IEEE 802.11e standard. 

The wireless systems can be configured in two different modes in the IEEE 802.11 architecture: (i) the ad-hoc mode, and (ii) the infrastructure mode. The infrastructure mode has multiple stations that can communicate with each other through an access point (AP), where the APs have connectivity with a wired network at the backend. While in the ad-hoc mode, the stations can communicate directly with each other without any intervening access points or a backend wired network. The IEEE 802.11 has two medium access coordination functions, namely the DCF and the PCF.
	
Various kinds of QoS enhancement techniques have been proposed for the IEEE 802.11 standard \cite{zhao2002performance} \cite{li2005} \cite{achary2012enhanced}, and they are explained in the rest of this subsection. 


%


\vspace{2mm}
\subsubsection{Priority Queueing}

This method is used to provide priority queues at the MAC layer where data packets are segregated on the basis of priorities. Whenever a particular station has access to the channel, it transmits the one which has the highest priority among the queued packets. All the stations must contend with each other for access to the medium. 

Priority queueing is done in a way that there are eight different levels of priority, and therefore eight different queues must be maintained. Table \ref{tab:priorityQueueing} shows the classification of these priorities queues. The highest level or the seventh level has the highest priority and it is assigned to the most critical applications. The next two levels, i.e. levels 5 and 6, correspond to delay-sensitive video and audio applications. Levels 4 and below are used for regular data traffic, as well as streaming video. Level 0 is left for the traffic that can tolerate all the deficiencies of the best-effort service \cite{sundareswaran2007improving}.

\begin{table}
\caption{Priority Levels Corresponding To Various Applications Types For Supporting Priority Queueing in IEEE 802.11.}
\scriptsize
\centering
\begin{tabular}{p{.8cm} p{1.4cm} p{1.6cm} p{3cm}}

\toprule
\textbf{\emph{Priority}} &  \textbf{\emph{802.1 D User Priority}} & \textbf{\emph{802.11e Access Category (AC)}} & \textbf{\emph{Description}}\\
 \midrule
Lowest & 1 & AC\_BK & Background Traffic \\
& 2 & AC\_BK & Background Traffic \\
& 0 & AC\_BE & Best Effort \\
& 3 & AC\_BE & Best Effort \\
& 4 & AC\_VI & Video \\
& 5 & AC\_VI & Video \\
& 6 & AC\_VO & Voice \\
Highest & 7 & AC\_VO & Voice, Network Management \\
 \bottomrule
\end{tabular}

\label{tab:priorityQueueing}

\end{table}

\vspace{2mm}
\subsubsection{Differentiated Services}

\begin{figure*}[t]
\centering
 \includegraphics[width=10cm]{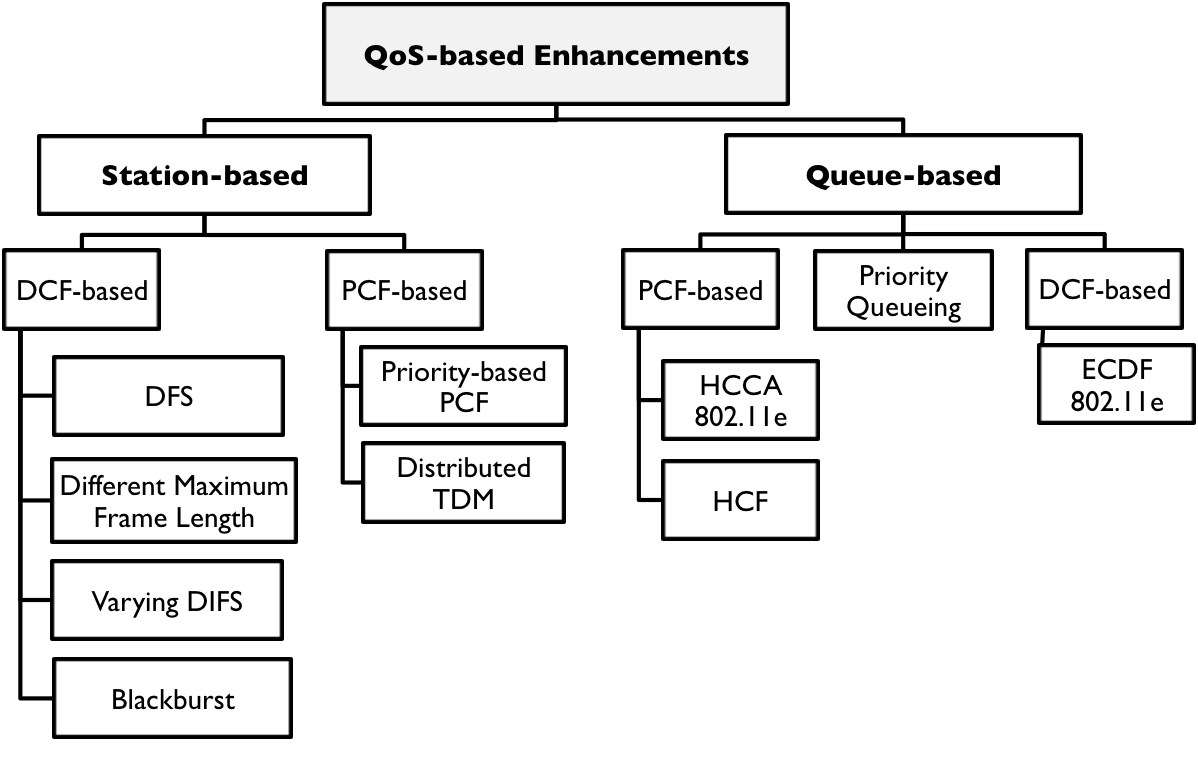}
\caption{MAC layer QoS enhancement schemes for IEEE 802.11-based wireless networks (described in Section \ref{sec:MAC}).}
\label{fig:qosmac}
\end{figure*}

The QoS enhancements can also be classified in the terms of the DCF-based or the PCF-based enhancements. Figure \ref{fig:qosmac} provides a taxonomy of DCF- and PCF-based enhancements \cite{ni2004qos} for both priority queueing and differentiated services.

\vspace{2mm}
We initially discuss four main techniques for deploying \textit{differentiated services using DCF}:


\begin{itemize}

\vspace{2mm}
\item \emph{Distributed Fair Scheduling}: For good performance of a system, it is not a fair practice to retrict the services of low-priority traffic and to provide better services to high priority traffic. One way is to assign more bandwidth to the high priority traffic in comparison to the low priority traffic. Distributed fair scheduling (DFS) is a technique used in this respect. In this technique, each flow is assigned some weight depending on its priority and the bandwidth it gets is then proportional to this weight. This is a centralized technique in the sense that it uses a central AP, which has the information regarding all the traffic flows from different stations, and can therefore assign different weights to each of these flows. This technique thus differentiates among all the traffic flows going through the AP \cite{lindgren2003quality}.

The DFS scheme uses the backoff mechanism of IEEE 802.11 to decide the transmission order of each station. When the transmission starts, each station chooses a random backoff time. This backoff interval is a function of packet length and the priority of the flow. The stations with low priority flows have longer backoff intervals than the stations with high priority flows. Using packet size in the backoff calculation ensures fairness amongst the stations, resulting in smaller packets being sent more often. In the case of a station experiencing a collision, the new backoff interval is generated using the same algorithm.

\vspace{2mm}
\item \emph{Varying DIFS}: Another solution is to vary the distributed inter-frame spacing (DIFS) duration for differentiation among flows \cite{aad2001differentiation}. For example, we know that the ACK packet in the IEEE 802.11 standard gets higher priority than RTS packets, due to the fact that ACK packet waits short inter-frame spacing (SIFS) amount of time, while RTS packet waits DIFS amount of time, which is much longer. The same idea can be taken to the data frames; in which each flow's priority is set with a different DIFS duration. To avoid collisions, a backoff time is maintained similarly in these packets as well. Such technique is much beneficial in real-time applications, where delays have a greater significance compared to packet loss \cite{drabu1999survey}.

\vspace{2mm}
\item \emph{Differentiated Maximum Frame Length}: In this approach, service differentiation is achieved in a way that different stations can transmit frames with different maximum frame sizes. The stations with high priority flows can transmit a larger frame than the one with the lower priority flows. To ensure this, there are two mechanisms: either the packets that exceed the maximum frame size are discarded or an upper bound on the size of packets is maintained in each station \cite{aad2000introducing}. In some cases, when the packet size is greater than the maximum limit, the packets are fragmented. These fragments are sent without any RTS in between, waiting just for the reception of corresponding ACKs. These mechanisms provide us with the same data rates as those without fragmentation \cite{{drabu1999survey}}.

\vspace{2mm}
\item \emph{Blackburst}: The blackburst scheme imposes certain constraints on high priority flows rather than the low priority flows which has been considered until now \cite{{sharma2013quality}}. In this technique, every station gets access to the medium for a fixed interval of time \cite{{wang2011mobility}}. Once the station gets access to the medium, it jams the medium for a certain duration. Consider a station that has higher priority than others, and it has data packets to transmit, so it senses the channel. Once it detects the channel has been idle for PIFS amount of time, it has the potential to transmit its frames. Hence, after waiting for a PIFS amount of time, it enters a blackburst contention period. A jamming signal, which is called blackburst, is then sent by this station to jam the channel. The length of this blackburst signal is proportional to the amount of time a particular station must wait before getting access to the medium. After the station has transmitted its blackburst signal, it again listens to check if any other stations are also sending a blackburst signal. The length of this blackburst signal is compared to check whether it is longer or shorter than its own. Subsequently, the station with the longest blackburst shows that it has been waiting for a longer amount of time to access the channel, hence it is the next station to access the channel. This technique is similar to how TDM shares the same medium among the different flows, and it is used in real-time traffic and synchronization \cite{{ni2004survey}}.

\end{itemize}

\vspace{2mm}
We next discuss two techniques for offering \textit{differentiated services using PCF}:

\begin{itemize}

\vspace{2mm}
\item \emph{Distributed TDM}: This mechanism uses a polling method as in the regular PCF mechanism, but time slots are also defined as in the TDM approach, and each of these time slots is assigned to a specific station. Once these time slots are assigned, each station knows when to transmit, and thus transmission of packets can be done with a very little involvement of the AP \cite{{drabu1999survey}}.

\vspace{2mm}
\item \emph{Hybrid Coordination Function}: Hybrid coordination function (HCF) is a new coordination function proposed in IEEE 802.11e to enhance both DCF and PCF. HCF uses two methods: the first method is contention-based and it is known as enhanced distributed channel access (EDCA), and the second method is contention-free and it is known as HCF-controlled channel access (HCCA). HCF uses the AP as a traffic manager which is termed as the hybrid coordinator (HC) \cite{{chen2011high}}, which is a centralized coordinator. The HC negotiates the exchange of frames and the frame handling rules given in HCF. The HC is located within the range of AP and works both in the contention-based and contention-free periods. The traffic is composed of wireless station (STA) ``streams'' or pipes, with each STA stream  associated with a set of QoS parameters \cite{{kowalski2013hybrid}} negotiated with the AP. The AP uses a polling method to control the traffic. It sends polling packets to the stations. When a station is polled, it replies to the poll in a frame that contains the response and the data to be transmitted. In this method, the polling is based upon the priority on which QoS has to be ensured \cite{{gargusing}}.

\end{itemize}

The various techniques for service differentiation covered in this section are summarized in Table \ref{tab:comparisonMAC} along with their main features and advantages.

\begin{table*}
\scriptsize
\renewcommand{\arraystretch}{2}
\caption{Comparison of the MAC layer service differentiation schemes using DCF in the IEEE 802.11 standard.}

\centering
\begin{tabular}{|p{3.2cm}|p{5.8cm}|p{5.6cm}|p{1.5cm}|}
\hline
\textbf{\textit{MAC scheme}} & \textbf{\textit{Main features}} & \textbf{\textit{Advantages}} & \textbf{\textit{References}}\\
\hline
Distributed Fair Scheduling (DFS)& The DFS algorithm uses the backoff mechanism as a function of packet length and the priority of the flow. & Provides fairness to all the flows; performance of high priority flows is increased. & \cite{lindgren2003quality}\\
\hline
Varying DIFS & Flow priority is given by setting different DIFS durations. To avoid collisions, a similar backoff time is maintained. & Provides benefits to real-time applications where higher delay is more significant than lower packet loss. & \cite{aad2001differentiation}\\
\hline
Differentiated maximum frame length & Services are differentiated by defining maximum frame size proportionately to a flow's priority. & Reduces contention overhead and achieves good differentiation. & \cite{aad2000introducing} \\
\hline
Blackburst & The blackburst scheme jams the channel according to the amount of time it has waited. & Minimizes delay of real-time flows; high priority flows get maximum benefit in the absence of low priority flows. & \cite{sharma2013quality} \cite{ni2004survey} \\
\hline
Enhanced Distributed Coordinated Function (EDCF) & EDCF is a contention-based channel access function of IEEE 802.11e which can provide differentiated service. & Provides better service differentiation using priority queues. & \cite{romdhani2003adaptive} \cite{qashi2011evaluating} \\
\hline
Hybrid Coordination Function (HCF) & The hybrid controller provides transmission opportunities to stations with higher priority packets. & Priorities are given based upon the channel conditions. & \cite{gargusing} \\
\hline
\end{tabular}
\label{tab:comparisonMAC}
\end{table*}

\vspace{2mm}
\subsubsection{QoS Scheduling}

A priority scheduler always selects packets from a queue with the highest priority. Such an approach is simple to understand, but can unfortunately lead to starvation of lower priority packets, particularly when there is a steady flow of high priority packets. There are also deadline-based and rate-based scheduling schemes.

The process of QoS scheduling in the IEEE 802.11 standard chooses packets amongst the various flows and distributes them on to specific links depending upon the requirements of each flow. This distribution of flows on each link has to be done within a small time interval and should be hardware-friendly. Scheduling is designed to provide a better throughput while reducing transmission times---throughput and delay being the key metrics quantifying better QoS. Resource reservation for different traffic flows requires synchronization among nodes to effectively monitor the changes in resource adaptation \cite{yu2013resource}. For such kind of insurances, we need to have a real-time monitoring mechanism for the changing network environment. The scheduler is generally operating at the MAC layer of the TCP/IP model.

In \cite{zhang2009qos}, a cross-layer design algorithm for QoS packet scheduling has been defined which considers delay and information shared at the PHY, MAC and network layers. It helps in high-speed data transmission through careful monitoring of the constant changes in the network while providing fairness to all flows. No that, for best-effort services, the scheduling scheme treats all flows with the same priority. Several QoS scheduling techniques have been proposed \cite{lu1999fair} \cite{tsao2000extending} and many enhancements of current techniques have been discussed \cite{grilo2003scheduling} \cite{lim2004qos} \cite{xiao2004ieee}.

In \cite{ansel2004efficient}, a HCF-based packet scheduler is designed and implemented with special reference to the IEEE 802.11e standard. This design caters both the constant bit rate and variable bit rate of the QoS-sensitive traffic and provides bandwidth support and smaller delays to all network flows. The QoS scheduling provides guaranteed services taking into consideration the bit rate, delay, throughput, etc. The general algorithms that ensure QoS in scheduling are: (i) strict priority; (ii) weighted fair queueing (WFQ); (iii) weighted round robin (WRR); and (iv) earliest due date (EDD).

\vspace{2mm}
\paragraph{Strict Priority}

In this algorithm, the buffer is partitioned into a number of different queues, which is equal to the number of different priority flows. The packets are then stored in these queues by the scheduler according to their own priority levels. The flows in the same queue are then sent using the FIFO scheme. The strict priority algorithm is easy to implement but it does not guarantee any bit rate and losses. Moreover, the lower priority flows may have a zero-valued throughput. In \cite{georges2004formal}, \cite{jasperneite2002deterministic}, and \cite{georges2005strict}, a network calculus method is used to evaluate the performance of a switch as it provides a good model of packet exchanges, and it determines end-to-end delay. Note that, the strict priority scheduling is implemented in Ethernet switches. A slight modification to the strict priority algorithm is proposed in \cite{jiang2002probabilistic}, where the different flows are assigned with different parameters. The technique is important in the per-hop behaviour of differentiated services network.

\vspace{2mm}
\paragraph{Weighted Fair Queueing}

The same idea of assigning each flow with a certain priority is used, however the queues are not served on FIFO. Each flow is assigned a specific weight according to the QoS requirements \cite{parekh1993generalized}. Hence, the bit rate varies with each flow. A certain upper bound on the buffer size is implemented to give all the flows a share of the bandwidth, which is unlike to what we have seen above. An interleaved WFQ scheme is implemented in \cite{chen2005interleaved}, where a table specifies the queue sequence. The table is interleaved, so higher priority flows are visited more frequently. The scheme improves on latency and jitter which are associated with the traffic queues. In \cite{banchs2002distributed}, the WFQ scheme that is backward compatible with the IEEE 802.11 standard is discussed. The simulation results show that the scheme can provide appropriate bandwidth distribution even in the presence of flows that need to be transmitted at all times.

\vspace{2mm}
\paragraph{Weighted Round Robin}

Weighted round robin is a frame-based implementation of WFQ. The flows are segregated similarly in separate queues with a specific weight assigned to each queue. The management can get difficult at times with different packet sizes. A new scheduling algorithm, called the dynamic WRR is proposed in \cite{kwon1998scheduling}. This algorithm is suitable for all traffic forms having variable and constant bit rates. The queues of traffic are assigned a dynamic weight. It helps the network in providing multimedia services even in the presence of bursty traffic. In \cite{kwak2002modified}, a modified dynamic WRR scheme is proposed. This scheme guarantees the delays in real-time traffic and provides efficient transmission of other forms of traffic.

\vspace{2mm}
\paragraph{Earliest Due Date}

In the normal EDD scheme for wired networks, packets of several different flows are assigned deadlines according to which packets are served first by the packet scheduler with the smaller deadline indicating higher priority. Since wireless networks show varying characteristics, the deployment of EDD is not an easy task. Therefore, in \cite{elsayed2006channel}, a channel-dependent EDD (CD-EDD) is described. It depends on the channel state, and the packets are queued by the scheduler on the basis of earliest expiry time and other channel parameters. The prioritized flow consequently gets the highest transmission rate among all the flows.

\vspace{2mm}
\subsubsection{Traffic Shaping}

Traffic shaping is used to control the flows of traffic in a channel. The basic idea is to limit the amount of packets per station. A traffic controller is used to comply the QoS requirements of each flow. Traffic shaping can split the resources according to different requirements of different flows. The traffic shaper must adapt to the variations in a channel. The traffic shaping mechanism has a strong impact on the performance of a system \cite{morris2008automatic}. Several traffic shaping parameters are used in the QoS model of IEEE 802.11 standard:  e.g., the aggregation level and the bursting level. \textit{Aggregation level} refers to the amount of packets that are aggregated into a single IEEE 802.11 packet. \textit{Bursting level} refers to the amount of packets transmitted at each transmission opportunity \cite{zhang2007qos}.

\subsection{QoS Support in IEEE 802.11e}
\label{sec:802.11e}

The IEEE 802.11e standard is an important extension of the IEEE 802.11 standard focusing on QoS \cite{mangold2003analysis} that works with any PHY implementation. Wireless nodes equipped with IEEE 802.11e features are now known as QoS stations (QSTAs) and they are associated with a QoS access point (QAP) to form a QoS basic service set (QBSS). The main feature of the IEEE 802.11e standard is that it improves the MAC layer for QoS provisioning by providing support for: (i) segregation of data packets based on priority requirements; (ii) negotiation of QoS parameters through a central coordinator or AP; and (iii) admission control. 

The IEEE 802.11e standard introduces a contention-based MAC layer scheme called extended DCF (EDCF) and a polling-based scheme called HCF controlled channel access (HCCA). Both these schemes are useful for QoS provisioning to support delay-sensitive voice and video applications \cite{choi2003ieee}, and they are described next.

\vspace{2mm}
\subsubsection{Extended DCF (EDCF)}

In the DCF configuration, a contention window is set after a frame is transmitted. This is done to avoid any collisions. The window defines the contention time of various stations who contend with each other for access to channel. However, each of the stations cannot seize the channel immediately, rather the MAC protocol uses a randomly chosen time period for each station after that channel has undergone transmission \cite{yang2002priority}.

EDCF uses this contention window to differentiate between high priority and low priority services \cite{romdhani2003adaptive}. The central coordinator assigns a contention window of shorter length to the stations with higher priority that helps them to transmit before the lower priority ones \cite{krithikaquality} \cite{qashi2011evaluating}. To differentiate further, inter-frame spacing (IFS) can be varied according to different traffic categories. Instead of using a DIFS as for the DCF traffic, a new inter-frame spacing called arbitration inter-frame spacing (AIFS) is used. The AIFS used for traffic has a duration of a few time slots longer than the DIFS duration. Therefore, a traffic category having smaller AIFS gets higher priority \cite{villalonprovisioning}.

\vspace{2mm}
\subsubsection{HCF Controlled Channel Access}

The HCF controlled channel access (HCCA) is IEEE 802.11e specific, and it makes use of a Hybrid Coordinator (HC) to manage the bandwidth allocation of wireless medium \cite{khanoptimization}. The HC can obtain a transmission opportunity (TXOP) and initiate data deliveries to provide transmission opportunities to a station with a higher priority without any backoff; that is to say, the HC can access the channels after a PIFS amount of time rather than a DIFS amount of time as for the other stations \cite{mangold2003analysis}. As PIFS is smaller than DIFS and AIFS, the HC has a priority over the DCF traffic, and also over the ECF traffic that uses AIFS.

\vspace{2mm}
\subsubsection{Control Access Period (CAP)}

The CAP in HCCA is a period when access to the wireless medium is controlled \cite{ni2005performance}. During this time, the HC, or the AP, gives the right of using the medium to a device. The AP can gain access to the medium before any other stations, and can then provide the transmission opportunity to any station. This guarantees data transfer from a station irrespective of the congestion level in the channel \cite{rashid2008controlled}. The AP can schedule such transmission opportunities for each of the stations, and can provide the parameters needed for QoS provisioning \cite{dujovne2010taxonomy}.

\vspace{2mm}
\paragraph{CAP in the Contention Period}

A CAP in the contention period is used to regulate access to the medium to guarantee various QoS parameters \cite{reddy2006quality}. However, in this method, the AP is not the controller (or the sole decision maker) \cite{gu2004sequential}. This is because any stations having the DCF traffic, or any EDCA traffic, can interfere with the scheduling performed at the AP, so this can delay the already scheduled data transfer at a particular station. Moreover, CAPs may use an RTS to prevent other devices while contending for the medium \cite{ni2005performance}. This causes a marginal overhead \cite{rashid2007hcca}.

\vspace{2mm}
\paragraph{CAP in the Contention Free Period}

The contention free period (CFP) is the most efficient way to use the channel, and it allows the AP to have a fine control of the medium \cite{cervello2006collision}. During this period, the AP has full command of the medium, so the stations do not contend for the access to the medium \cite{yeh2002support}. The scheduling of the traffic, and the provision of QoS guarantee to the stations, is handled by the AP itself. The AP can set multiple CAPs following each other and uses the smallest possible time intervals to separate every CAP \cite{rashid2007hcca}.

\vspace{1mm}
\subsubsection{Other IEEE 802.11e QoS features}

We now outline some other important features of the IEEE 802.11e standard. The TXOP parameter defines a time limit for the utilization of radio resources at the stations \cite{mangold2003analysis}. The automatic power save delivery (APSD) mechanism is used by the AP to deliver multiple frames within a service period. Thereby, APs can enter sleep period until the next service period to conserve energy \cite{perez2010ieee}. The APSD mechanism has scheduled APSD and unscheduled APSD, both of which lead to power saving as compared to the legacy IEEE 802.11. The IEEE 802.11e standard also supports block ACKs for the acknowledgment of multiple MAC protocol data units (MPDUs) in a single block acknowledgment frame resulting in reduced overhead \cite{tinnirello2005efficiency}. 

The NoAck is another enhancement that indicates the loss of a packet, so retransmission can be ensured quickly to reduce delay \cite{politis2011exploiting}. Direct link setup is another supported feature that allows direct station-to-station transfer within a service set.


\section{Network Layer QoS Solutions for IEEE 802.11}
\label{sec:NET}

The bulk of research investigating QoS solutions for the network layer of IEEE 802.11 networks has focused on admission control and QoS routing. These two important facets of network layer QoS solutions are discussed next in separate subsections. 

\subsection{Admission Control}

Although the enhancements explained at the MAC layer provides service differentiation among different traffic flows, it can ensure QoS only when network load is reasonable. If the load increases beyond a certain limit, the QoS guarantees are not ensured even to high priority traffic \cite{mangold2002ieee} \cite{xiao2004local}. This is where the admission control mechanism helps in preventing the network from becoming congested, by allowing or disallowing flows depending on whether the conditions are favorable to meet QoS requirements. More specifically, the purpose of admission control is to limit the amount of newly admitted traffic such that the QoS performance of existing flows is not degraded \cite{gao2005admission}. Admission control is a key component to adapt to the traffic variations according to the changing environment of IEEE 802.11-based wireless networks \cite{andreadistechniques}. In \cite{hanzo2009admission}, Hanzo et al. has presented a very comprehensive survey on different admission control schemes available in the literature. Admission control can be categorized into three different methodologies \cite{brewer2010comparison}.

\vspace{2mm}
\subsubsection{Measurement-based Admission Control}

In this scheme, the decisions are made through continuous monitoring of network status, such as throughput and delay. A certain threshold is maintained according to the network status for admission of new traffic flows. Nor et al. in \cite{nor2006admission} proposed a metric called network utilization characteristic (NUC) as a means for admission of traffic flows into network. NUC defines the amount of channel utilized to transmit the flow over the network. This scheme guarantees QoS to high priority flows under loaded channel environments. Another scheme presented by Wu et al. in \cite{wu2010design} is that each traffic class is assigned a certain portion of available resources, and these resources are then remaining reserved for that particular class. In this regard, only the traffic with higher priority compared to the existing traffic is admitted.

\vspace{2mm}
\subsubsection{Model-based Admission Control}

In model-based schemes, the network status is measured based on some models. The Markov chain models are quite popular in attempts at modeling IEEE 802.11 although other approaches are also being explored due to some limitations of Markovian models \cite{chen2006supporting}. In \cite{cano2007adaptive}, an analytical model is used to estimate the minimum bandwidth requirement of all flows. When a newly admitted flow need to be activated, the algorithm checks if it is going to result in preservation of QoS requirements of existing flows.

\vspace{2mm}
\subsubsection{Measurement-aided, Model-based Admission Control}

It is a hybrid of measurement-based and model-based schemes. The algorithm in \cite{ksentini2007etxop} takes network measurements in a loaded environment and also the data rate requirements of the flow that is requesting for admission. Furthermore, a channel model is applied to predict the network conditions and provides QoS enhancements accordingly. Another solution is the threshold-based approach proposed in \cite{bensaou2009measurement} in which the channel conditions are continuously monitored and the contention probability is measured. When any new flows request for admission, the admission control checks for the competing flows. The absolute bandwidth and the expected delay of the new flow are measured. If this satisfies the threshold conditions, then this flow is admitted.

\subsection{QoS Routing Schemes}


QoS routing is an essential part of the overall QoS architecture in the IEEE 802.11 standard. QoS routing allows the network to compute a path that supports the QoS objectives of various flows under the constraints of wireless medium. The chosen path may or may not be the shortest path, but it meets a particular service category objectives \cite{sivakumar1999cedar} \cite{yin2006traffic}. As an example, Matos et al. proposed to compute routing decisions of voice, video, and data in a decentralized fashion at intermediate nodes in wireless multi-service networks such that the overall network performance is optimized per the desired QoS \cite{matos2012quality}.

There are various metrics that can be used for measuring QoS routing performance. We describe an example work for each metric. The metrics proposed for QoS-based routing in the literature are: (i) minimum throughput, or capacity, required in bits per second \cite{lin1999qos}; ii) maximum tolerable delay in seconds \cite{chen1999distributed}; iii) maximum tolerable packet loss ratio (PLR) \cite{abdrabou2006position}, and iv) maximum tolerable jitter \cite{bashandy2005generalized} \cite{wang2005application}. In addition to these generic metrics, there are also other metrics specific to various layers of TCP/IP model. For example, at the \emph{network layer}, achievable throughput or residual capacity \cite{lin1999qos}, end-to-end delay \cite{chou2006low} \cite{chen1999distributed}, node buffer space \cite{sheng2003routing}, and route lifetime \cite{rubin2003link} are important metrics; at the link (or MAC) layer, link reliability \cite{barolli2003qos}, and link stability \cite{rubin2003link} are important performance metrics; finally, signal-to-interference ratio (SIR) \cite{kim2004demand}, bit error rate (BER) \cite{wisitpongphan2005qos}, and node's residual energy \cite{toh2001maximum} are important performance metrics at the PHY layer. A comprehensive summary of QoS-based routing metrics is provided in a survey paper \cite{hanzo2007survey}.
 
There are a number of frameworks that can be used for QoS-based routing. We describe two such frameworks. 

\vspace{2mm}
\subsubsection{Measurement-based QoS Routing}

In \cite{liu2004courtesy}, a framework is provided to achieve fairness among different priority flows. The basic concept is to allow the high priority traffic to help the low priority traffic by sharing their unused bandwidth. This scheme can considerably improve the system performance and it can shorten the delays when the traffic load is very high. The QoS-supporting algorithm presented in \cite{chen2005qos} helps applications to find routes that satisfy their service needs, or a feedback is provided in case of non-availability of these resources. Thus, a protocol that is QoS-aware and also has admission control and feedback mechanism is proposed. Another resource reservation algorithm is proposed by Xue in \cite{xue2003ad} in which bandwidth and delay are measured very accurately using a MAC protocol with collision detection. These calculations are then used by the algorithm to make decision in admission and reservation of resources.

\vspace{2mm}
\subsubsection{Ticket-based Probing Algorithm}

This algorithm uses tickets to limit the number of paths observed. When any source wants to get a QoS satisfying path to any destination, it sends probe messages along with a ticket. The number of tickets is equivalent to the number of paths searched. When the destination receives this probe message, the path from source to destination is set \cite{chen1999distributed}.

\section{Transport layer QoS Solutions for IEEE 802.11 }
\label{sec:Transport}

The classical version of TCP protocol performs rate control based on its assumption that packet losses occur solely due to network congestion. This assumption does not hold true for wireless networks where channel noise and interference can be another significant cause of packet loss. With such an assumption, TCP performs poorly in terms of end-to-end QoS since it may reduce the sending rate even in uncongested networks under the mistaken assumption that packet losses are only caused by network congestion.

The problem of suboptimal performance of TCP in wireless networks has been known for long and much effort has focused on improving TCP's performance \cite{{balakrishnan1997comparison}} \cite{{chen2002syndrome}}.
Most of the existing work in the IEEE 802.11 standard is focused on the QoS requirements of multimedia applications, such as VoIP \cite{jelassi2012quality}, and data traffic, such as web, email, media downloads, etc. But with growing demand of wireless networks, time-critical applications with voice and video do place significant QoS requirements on wireless medium. To upgrade QoS at the transport layer, we can adopt several techniques taking into account the delay and loss as the basic parameters. With the main focus on congestion control and error control, several techniques are discussed below.

\subsection{Congestion Control}

The bursty nature of the wireless media and path loss causes the degradation of services to applications that require high video quality. This degradation is caused by network congestion. This is why TCP congestion control is essential to minimize packet loss and reduce delay. Rate control is a congestion mechanism which reduces network congestion by comparing the required bandwidth for video with the available bandwidth \cite{{bolot1998experience}}. Multiple standards of rate-adaptive video encodings \cite{{wu2000end}} exist for different applications, such as H.261 and H.263 for video conferencing \cite{{martins1996joint}} \cite{{wiegand1996rate}}, as well as MPEG-1 and MPEG-2 for real-time transmission \cite{{ding1997joint}} \cite{{hsu1997joint}}. The main purpose of a rate-adaptive encoding scheme is to enhance the video quality under a certain encoding rate. Rate control and rate shaping \cite{{ding1997joint}} are the algorithms for congestion control in the IEEE 802.11 standard.

\subsubsection{Rate Control}

It is very important for an end-to-end protocol to accurately estimate the appropriate sending rate for network transfer since an infeasibly high sending rate can result in packet losses and retransmissions. TCP retransmissions that result from packet losses may lead to unacceptably long delay for QoS-aware multimedia delivery over the wireless channels. Rate control subsumes flow control and congestion control which adjust sending rates to ensure that the sender's rate does not overwhelm the receiver and the network, respectively. 

Two types of congestion control are in wide practice: window-based \cite{{jacobson1988congestion}} and rate-based \cite{{turletti1996videoconferencing}}. The window-based approach analyzes the available network bandwidth by gradually increasing the size of congestion window. When congestion is detected (through the detection of packet loss), the protocol decreases the window size by a large amount. The abrupt decrease in the window size in response to congestion is necessary to prevent network failure. Window-based control performs retransmissions which result in extensive delay, which is intolerable in case of real-time video transmission. The rate-based control approach sends at a rate based on an estimated available network bandwidth. If the estimated bandwidth is accurate, then network congestion can be avoided. The rate-based control approach is usually used for transportation of real-time video. Existing rate control mechanism for real-time video are source-based, receiver-based or hybrid. The interested reader is referred to the paper \cite{zhu2011intelligent}, and the references therein, for a description of rate-control work focusing on supporting real-time traffic in WLANs.

\subsubsection{Rate Shaping}

Rate shaping is the practice in which the compressed video bit stream adjusts itself to the rate of a target rate. We can think of a rate shaper as an interaction medium between an encoder and the network, which matches the encoder's output to the available network bandwidth. Since rate shaper does not need any interaction with the encoder, it can be used for any video coding scheme for both live and stored videos.

\subsection{Error Control}

The QoS guarantees can also be made through error control. The main role of congestion control is to avoid packet loss. However, we are unable to avoid packet loss completely in the Internet, and as a result the quality of video or other bandwidth hungry applications are affected. The error control schemes at the transport layer are application-aware. The error recovery schemes can be divided into two basic types:

\subsubsection{Automatic Repeat Request (ARQ)}
The ARQ scheme uses an acknowledgement packet to indicate that a packet has been received successfully. It is very efficient for high-speed wireless links because the round trip delay of the link is very small \cite{{nam2002experimental}}. The ARQ scheme can be implemented at both transport and link layers of the OSI model \cite{{crow1997ieee}}. The traffic is segmented into queues such that QoS guarantees are ensured. The ARQ scheme can adapt to channel errors and is more efficient in terms of bandwidth utilization.

\subsubsection{Forward Error Correction (FEC)}
The FEC scheme adds redundant bits to the flow which helps in recovering the erroneous bits. The FEC is used for the transmission of real-time applications which have a strict delay requirements \cite{{liu1997error}} \cite{{aikawa1996forward}}. However, a drawback of FEC is the increased overhead even in the absence of errors in the link \cite{{choi2006ieee}}. FEC helps in maintaining a uniform throughput and time delay in the networks. However, the overhead increases with channel errors because long FEC codes must be used.


\subsection{Prioritization}

The TCP ACK prioritization method uses both the AIFS and the minimum contention window (CWmin) parameters. The stations having smaller CWmin gets more transmission opportunities than stations having larger value of CWmin as their backoff counter is smaller. The AIFS parameters can be used to allow the AP to have quicker access to the wireless medium. Since the TCP ACKs can go freely through the bottleneck links, the performance of the system is upgraded \cite{{leith2005tcp}}.

\section{Application Layer QoS Solutions for IEEE 802.11 }
\label{sec:Application}

Apart from work at the lower layers, it is also possible to implement QoS in IEEE 802.11 networks at the application layer. Traditionally, the work done on application layer QoS has focused on various aspects of multimedia delivery using techniques such as scalable video coding \cite{van2006optimized}, error correction coding and rate-distortion optimization \cite{chakareski2003rate}, source coding \cite{luo2006optimal}, transcoding \cite{luo2008video}, adaptive transmission \cite{luo2008video}, as well as rate control \cite{luo2006optimal}. We note here that most of the QoS work done at the application layer is cross-layered in nature, specifically drawing upon information from, and interacting with, other layers. We provide two brief examples as illustration.

Chakareski et al. \cite{chakareski2003rate} proposed an optimization-based \emph{error-correction coding} scheme, which works at the application layer, for rate-distortion optimized multimedia streaming to wireless clients. The scheme allows the sender to compute the prioritization levels of packets to satisfy an average transmission rate constraint while minimizing the average end-to-end distortion 

Van der Schaar et al. \cite{van2006optimized} proposed a framework for delay-constrained video streaming over IEEE 802.11 a/e WLANs. The authors considered the problem of video transmission over HCCA and developed a cross-layered optimization framework working at the PHY, MAC, and application layer of the TCP/IP layered model. In another cross-layered video streaming paper \cite{li2004providing}, Li and Van Der Schaar proposed an error protection scheme for transmission of layered coded video to provide adaptive QoS through prioritized queuing at the network layer and limitation of retries (or retransmissions) at the link layer. The basic insight of this work is that different video layers of varying importance may not receive uniform processing and protection, but may receive unequal priority depending on the channel conditions.


 The techniques for enhancements of various layers (e.g., MAC, network, transport and application layers) are summarized in Table \ref{tab:layeredoverview}.

\begin{table*}
\caption{Layered Overview of QoS Techniques in the IEEE 802.11-based Wireless Networks.}
\scriptsize
\centering
\begin{tabular}{p{3.3cm} p{11cm} p{2.5cm}}
\toprule
 \textbf{\textbf{\emph{QoS Enhancement Technique(s)}}} & \textbf{\emph{Description}} & \textbf{\emph{Reference(s)}} \\
 \midrule

 \vspace{1mm}

 \textbf{\emph{\underline{MAC Layer}}}\\ 
 
Priority Queueing & Data packets are segregated based on their priorities in queues. The packets with the highest priority are transmitted first and so on. & \cite{sundareswaran2007improving} \cite{ni2004qos} \\

Distributed Fair Scheduling & Each flow is assigned bandwidth according to its priority. & \cite{lindgren2003quality} \\

Varying DIFS & DIFS is varied in order to differentiate among flows. Each flow's priority is set by giving it a different DIFS. \\

Maximum Frame Length & High priority stations can transmit larger frames comparatively. & \cite{drabu1999survey} \\

Extended DCF & Shorter contention windows are assigned to higher priority stations helping them to transmit first. & \cite{krithikaquality} \cite{villalonprovisioning} \\

Blackburst & Contention period is used to indicate the waiting time for medium access. & \cite{ni2004survey} \\

 
 \vspace{1mm}
 \textbf{\emph{\underline{Network Layer}}} \\
 
Admission Control & The network is thoroughly examined and when congestion occurs, the nodes decrease their best-effort traffic in response. & \cite{domingo2004interaction} \cite{mangold2002ieee}\\

QoS Routing & Some routing mechanisms are used under which QoS paths are determined. QoS path may or may not be similar to the shortest path. & \cite{zhang2005qos} \cite{sivakumar1999cedar} \cite{yin2006traffic} \cite{matos2012quality} \\

 \vspace{1mm}
\textbf{\emph{\underline{Transport Layer}}}\\ 
 
Congestion Control & The congestion control mechanism orders the source to transmit traffic at a rate that is not greater than the available network bandwidth. & \cite{bolot1998experience} \cite{{jacobson1988congestion}} \cite{{turletti1996videoconferencing}} \\

Error Correction & Forward Error Correction and Automatic Repeat Request are used to ensure reliability. & \cite{nam2002experimental} \cite{aikawa1996forward} \cite{crow1997ieee} \\

TCP ACK Prioritization & The Contention Window and AIFS are used for flow prioritization. & \cite{leith2005tcp} \\

\\

 \vspace{1mm}
\emph{\underline{\textbf{Application Layer:}}}\\

Scalable Video Coding &  A cross-layered framework is proposed for delay-constrained video streaming over IEEE 802.11 a/e WLANs. The framework works at the PHY, MAC, and application layers of the TCP/IP layered model. &\cite{van2006optimized} \\
Transcoding & Video transcoding reencodes the stream to adapt the bit rate to the available resource. & \cite{van2004adaptive}\\
Application layer Error Control & An application layer optimization-based error correction coding scheme for rate-distortion optimized multimedia streaming to wireless clients. & \cite{chakareski2004application}\\
Real-time Retry Limit Adaptation & A real-time retry-limit adaptation is proposed at the link layer for video with adaptive QoS. & \cite{li2004providing}\\
Hybrid ARQ/ FEC & Multicast and unicast real-time video streaming approaches over WLANs are implemented through a hybrid ARQ algorithm that combines FEC and ARQ. & \cite{majumda2002multicast}\\

\\
 \bottomrule
\end{tabular}
\label{tab:layeredoverview}
\end{table*}

\section{Cross-layer QoS Solutions for IEEE 802.11}
\label{sec:CrossLayer}

While most QoS enhancement techniques are implemented at the MAC layer, wireless QoS can benefit from cross-layered interaction and implementation \cite{{toumpis2003performance}}. Since QoS provisioning entails various issues that span the range of the TCP/IP layered stack, cross-layer solutions are finding increasing deployment.

This section discusses a few cross-layered solutions for implementing QoS in wireless networks in general, and in IEEE 802.11 networks in particular. The various techniques for cross-layer enhancements discussed in this section are summarized in Table \ref{tab:Crosslayer}.

\begin{table*}

\caption{Overview of Cross-layer Enhancement Techniques in the IEEE 802.11-based Wireless Networks.}
\scriptsize
\centering
\begin{tabular}{p{4.5cm} p{10.5cm} p{2.0cm}}

\toprule
 \textbf{\emph{Feature(s)}} & \textbf{\emph{Description}} & \textbf{\emph{References}}\\
 \midrule
  Wireless Multimedia & SISO is used with adaptive modulation schemes at the PHY layer; and the impact on delay bound is measured at the data link layer. & \cite{{indumathi2010adaptive}} \\
\\
  Adaptive Modulating Scheme & The node with the highest priority is assigned resources first, and each type of connection adopts AMC at the PHY layer. & \cite{{liu2006cross}} \\
\\
  Wireless Scheduling & The information received from the PHY layer is utilized, and an efficient cross-layer packet scheduling approach is proposed which guarantees QoS parameters like delay, BER and received signal strength. & \cite{{abd2006efficient}} \\
\\
  Cooperative Communications & QoS is ensured by using cross-layer design taking into account the PHY and network layers using cooperative communication. & \cite{{sheng2011cooperative}} \\
\\
  Channel Coding and Retransmissions & The use of hybrid schemes corrects the errors in an adaptive manner by using a combined scheme of channel coding and retransmissions, thus improving the TCP performance. & \cite{{girod1999feedback}} \cite{{farber1999analysis}} \\
\\
  SoftMAC & SoftMAC regulates real-time and best-effort services in the network. & \cite{{wu2007softmac}} \\
\\
  OSAR and OMAR & Both are used in opportunistic networks for scheduling and
routing. & \cite{{zhang2008cross}} \cite{{wang2006omar}} \\
\\
  Dynamic Priority Functions & Defined for new nodes which are updated dynamically. AMC and MIMO are used accordingly for QoS provisioning. & \cite{{liu2006cross}} \\


Cross-layer Perceptual ARQ & A cross-layer priority-based ARQ algorithm for H.264 video streaming in IEEE 802.11 wireless networks. & \cite{bucciol2004cross}\\

\\
 \bottomrule
\end{tabular}
\label{tab:Crosslayer}
\end{table*}

\subsection{Cross-layer Features for Wireless Multimedia}

With increasing multimedia traffic on the Internet and wireless access being anticipated to become the future predominant Internet technology \cite{zander2013riding}, delivering multimedia applications with enhanced QoS has  become extremely important. Wireless channel information cannot be predicted easily due to deep fades and multipath effects, but getting information of source motion using video sequence is not hard. Due to the error-prone nature of wireless medium, and the undifferentiated nature of Internet's best-effort service model, multimedia delivery over the wireless networks is technically challenging. To improve user experience over the wireless Internet, QoS support can be introduced at different layers. 

In \cite{zhang2005cross}, the authors propose a cross-layered architecture---combining application-level, transport-layer, as well as link-layer controls---for supporting multimedia delivery over wireless Internet. These controls incorporate issues such as dynamic estimation of network and channel; adaptive error control, congestion control, and ARQ mechanisms; and priority based scheduling. In another work \cite{{qu2006source}}, a cross-layer framework is proposed in which the source motion is captured from a video sequence, and it consists of a packetization scheme, a cross-layer FEC-based unequal error protection scheme, and an intra coding rate selection scheme. This significantly improves transmission of bursty traffic and its losses over the wireless network without making the system complex.

In order to ensure QoS in real-time applications, we can bound delay instead of high spectral efficiency \cite{{indumathi2010adaptive}}. For example, we can ensure QoS for multimedia applications by analyzing the impact of the PHY layer on the data link layer. The single input and single output (SISO) mechanism is used with adaptive modulating schemes at the physical layer; and at the data link layer, we check the impact of the physical layer on the delay bound. Physical layer is modeled using a finite-state Markov chain. The use of appropriate scheduling schemes, and the resources allocated to the users can thus ensure high QoS for each user. This technique allocates resources to real-time users in time slots in a dynamic method using SISO along with adaptive modulating codes. 

\vspace{2mm}
\subsection{Adaptive Modulating Scheme (AMC)}

There are various works in literature that have proposed combining QoS reservation and scheduling at the MAC layer with adaptive modulation and coding (AMC) at the PHY layer. For instance, Liu et al. \cite{liu2005cross} have proposed an hybrid architecture combining QoS reservation and scheduling at the MAC layer with AMC at the PHY layer. With AMC, the physical transmission parameters can adapt to the changes in the link quality. As an example, the PHY layer can fallback to a modulation scheme more robust to noise, such as BPSK instead of QAM-16, in the case link quality degrades. 


In \cite{{agarwal2013optimal}}, the authors derive an optimal policy to reduce the average amount of dropped packets of a delay-controlled wireless node. The presented framework utilizes adaptive modulation for transmission of the optimal amount of packets to satisfy the QoS requirements. This framework, devised as a Markov decision process (MDP), works on reducing the long-term packet drop rate. 
In another work \cite{{liu2006cross}}, a cross-layered approach is presented for mobile wireless networks which studies the impact of the PHY layer infrastructure over the link layer QoS performance. This work considers MIMO diversity schemes along with AMC in its PHY layer analysis, while also studying the impact of the PHY layer infrastructure on real-time multimedia QoS provisioning performance at the link layer. 

\vspace{2mm}
\subsection{Wireless Scheduling}

The interaction between packet scheduling and the PHY layer is studied in \cite{{abd2006efficient}} where the network is used efficiently by predicting the future state of the wireless channel, as well as controlling the transmission power in multipath fading wireless CDMA networks. It is based on cross-layered model in which the information received from PHY layer is utilized by the scheduler, and an efficient cross-layer packet scheduling is proposed which guarantees users guaranteed QoS performance in terms of delay, BER, and received signal strength.

\subsection{Cooperative Communications}

QoS can also be ensured by using cross-layered design taking into account the PHY and networking layers using cooperative communication \cite{{sheng2011cooperative}}. Cooperative communication is first studied at PHY layer, followed by routing to ensure QoS in the network: i.e., we use an optimized link cost for the decision making of our routing leading to better path selection. The power consumption is kept to a minimum possible value, and end-to-end reliability is achieved by reducing the error rate. The selection of the best possible path leads to end-to-end reliability and thus the ensuring of QoS. In \cite{{xianyang2014design}}, the authors describe a novel concept named cooperative QoS routing, which sets up a routing path that helps to satisfy user's bandwidth requirement. The authors propose an optimization problem, called `widest cooperative routing path'(WCRP) problem, which finds a cooperative routing path with the maximum uninterrupted bandwidth and a scheduling scheme to evade interference.

\subsection{Combining Channel Coding and Retransmissions}
The transport layer ensures a reliable transmission by re-sending corrupted packets due to congestion. However, packets may get corrupted in wireless networks due to other reasons such as fading and multipath effects. The requests for repeated transmission for the packet in such cases would negatively impact the performance of the network. To cater to this problem, joint cross-layer techniques are used. FEC and ARQ are used as error correction codes; where FEC is used in delay-sensitive applications \cite{{boutremans2003adaptive}}, while ARQ is used for delay-tolerant applications (e.g., audio/video streaming). In most of the applications, FEC does not negotiate with the receiver for error correction because of the delay-sensitive nature although feedback from the receiver can be effective \cite{{girod1999feedback}} \cite{{farber1999analysis}}. Using hybrid schemes corrects the errors in an adaptive manner by using combined channel coding and retransmissions thus improving the TCP performance. Recently, network coding has been extensively used in wireless networks to upgrade the limited wireless capacity. In WLANs, network coding can be applied to packet retransmission. More than one packet can be evenly transmitted by a single retransmission at base station. In \cite{{tanigawa2011qos}}, the retransmission is based on network coding, and it cooperates with IEEE 802.11e EDCA. Consequently, QoS of high priority group is upgraded from the aspect of efficient loss recovery.

\subsection{Joint Cross-layer Techniques}
This subsection presents four main joint cross-layer techniques.

\vspace{2mm}
\subsubsection{Joint Rate Control, Admission Control, and Scheduling}

The main goal of an Internet designer is to share the resources efficiently. To use the scarce bandwidth in an efficient way, it is usually the case that real-time traffic co-exists with the rest of the traffic. We can jointly solve rate-control, admission-control, and scheduling problems for optimized solutions. In infrastructure-based networks, the eDCF is providing QoS to the nodes; but in distributed multihop networks, it is not possible due hidden terminal and interference problems. Thus, admission control puts a limit on real-time traffic that can overwhelm the system. By collaboration with nearby nodes, the interfering best-effort traffic is cut down to reduce the contention of real-time traffic. A scheme called SoftMAC is proposed in \cite{{wu2007softmac}} to use a control mechanism in order to regulate real-time and best-effort traffic in a distributed manner by coordinating with neighboring nodes. SoftMAC uses admission control to make sure that real-time traffic has sufficient bandwidth along its path. It also caters for rate control to avoid collisions between the real-time traffic and the best-effort traffic. Moreover, it has a priority queueing module to prioritize the real-time traffic.

QoS for wireless networks can be ensured by joint solutions working at the PHY and network layers. Such solutions can include joint routing and rate allocation to ensure QoS for different applications. Zhang et al. have proposed a framework for cross-layer design for QoS support in multihop wireless networks, and have reviewed in detail the interplay between joint routing at the network layer and rate allocation at the transport layer. In another paper \cite{{zhu2007rate}}, Zhu et al. discuss an optimization of joint allocation rate and multipath routing which allocates rates depending upon the distortion rate and congestion level. This paper considered the problem of rate allocation for multi-user video streaming sharing multiple  heterogeneous access networks. The problem was formulated as a convex optimization problem and distributed approximation of the optimization was proposed. 

\vspace{2mm}
\subsubsection{Joint Power Control, Scheduling, and Routing}

Various works in literature have demonstrated the deficiencies of the traditional siloed approach of independently performed power control, scheduling, and routing in wireless networking \cite{zhang2008cross} \cite{toumpis2003performance}. It is worth highlighting the strongly coupled nature of the these problems by noting that a change in power allocation, or the schedules on a given link, can impact flows that do not utilize the modified link. Due to the strong coupling between the network, MAC, and PHY layers, the power control, scheduling, and routing problems are best addressed jointly. 

With joint scheduling and power control, a network can generally achieve higher throughput and lower delay in a network \cite{zhang2008cross}, although for some unbalanced topologies, scheduling alone cannot satisfy bandwidth requirements and rerouting is also needed to send some packets through alternative routes and thereby release congestion. The routes are then selected according to the joint metric of energy consumed and traffic accumulated, with priority given to nodes with longer queue length. A similar approach is presented in \cite{{cruz2003optimal}} which aims to minimize the total average transmission power in a wireless multi-hop network through optimal link scheduling and power control. This work, however, requires tight time synchronization between transmitters, and quasi-static channel conditions that remain constant over several time slots, thus limiting the applicability of this work to interconnecting stationary nodes. 


\vspace{2mm}
\subsubsection{Joint Scheduling and Rate Optimization for Opportunistic Transmission}

In order to utilize the scarce resources of the wireless networks, opportunistic transmission takes advantage of the varying nature of the channel and improves throughput of the network. Two approaches are used in this regard: the first one uses the time diversity of an individual link by changing the transmission rate according to the channel conditions \cite{{moh2009link}} \cite{{liu2003opportunistic}}, while the second one considers multi-user diversity and jointly considers the time and spatial heterogeneity of a channel.   
Wang et al. \cite{wang2004opportunistic} have proposed a MAC solution named `opportunistic scheduling and auto rate' (OSAR) which jointly considers rate adaptation and multi-user diversity. OSAR protocols exploits the channel variations  by automatically adjusting the sending rate to best match the channel conditions. In a followup work, Wang et al. have proposed another solution `opportunistic medium access and adaptive rates' (OMAR)  \cite{wang2006omar}, which aims at efficient utilization of the shared medium in IEEE 802.11-based ad-hoc networks through joint consideration of multi-user diversity, distributed scheduling, and adaptivity. OMAR uses a clustering framework in which a node with a predefined number of links can function as the clusterhead to locally coordinate multiuser communication. The clusterhead is responsible for initiating medium access, while the cluster embers make medium access decisions in a distributed manner. 

\vspace{2mm}
\subsubsection{Joint Channel Assignment and Routing}

In this section, we present techniques considering the data link layer and network layer jointly. Interference among the channels is one of the main hurdle in achieving QoS in wireless networks. Orthogonal channel assignment is a potential solution to this problem. In \cite{{kyasanur2006routing}}, the authors present a joint algorithm for channel assignment and routing. The channel assignment algorithm performs two functionalities, the first one assigns channels on the basis of network topology, and the second function is to deal with the traffic information and assigns channels accordingly. Similarly it caters for creating multiple routes in the network to achieve higher throughput.

\section{Modern Network Architectural Trends and Wireless QoS}
\label{sec:modernArch}

This section presents three types of modern network architectures. Various research areas of recent architectures discussed in this section are summarized in Table \ref{tab:example}.

\begin{table*}
\caption{Sample of Research on QoS in IEEE 802.11-based Wireless Networks With Recent Architectural Developments}
\scriptsize
\centering
\begin{tabular}{p{5cm} p{11cm} p{1cm}}
\toprule

\textbf{\emph{Project}} & \textbf{\emph{Description of QoS Enhancement Technique(s)}} & \textbf{\emph{Reference(s)}} \\
 \midrule

 \vspace{1mm}
 \textbf{\emph{\underline{Software Defined Wireless Networks}}}\\
  
meSDN & Achieves real-time detection of QoS demands in a network and provides end-to-end QoS control. & \cite{lee2014mesdn} \\

QoSFlow & Provides packet scheduling algorithm to improve QoS mechanism in OpenFlow/ SDN-based networks. & \cite{{ishimori2013control}} \\
 
OpenQoS & Provides a dynamic routing scheme that generates shortest path for data delivery in order to minimize packet loss and latency. & \cite{{egilmez2012koc}} \\
  
OpenFlow Controller for Multimedia Delivery & Facilitates multimedia delivery with QoS using the best path with optimal service configuration.& \cite{{kassler2012towards}} \\
  
FlowVisor & Supports `QoS-enabled network slicing' that provides a user or an application with a certain network capacity ``slice'', which is isolated from other coexisting slices servicing other users/networks.& \cite{sherwood2009flowvisor} \\
  
Multimedia Streaming QoS Architectures for SDN & Provides QoS extensions for multimedia delivery using distributed control architecture in multi-operator SDNs. & \cite{egilmezdistributed} \\

Interference Mitigation in Enterprise WLAN &  Proposees an OpenFlow-based framework for interference mitigation in enterprise WLANs using SDN/OpenFlow. &\cite{zhao2014leveraging} \\

OpenFlow-based QoS support for Ofelia & Proposes architectural extensions to make Ofelia a QoS-supporting federated experimental testbed. & \cite{sonkoly2012qos}\\

OpenQFlow & Proposes a flexible variant of OpenFlow supporting a two-tiered flow-based QoS framework. & \cite{nam2013openqflow}\\
  
\vspace{1mm}
 \textbf{\emph{\underline{Cloud-based Wireless Networks}}} \\
 
EDCA model for Cloud & Proposes an EDCA model for QoS-aware differentiated multimedia cloud service provisioning in WLAN networks. & \cite{tursunova2012realistic} \\

Resource Allocation in Clouds & Proposes energy-aware resource allocation mechanisms for data centers set up in cloud environments. & \cite{{sharkh2013resource}} \\

Dalvi et al. & Proposes centralized cloud-based approaches for managing WLANs. & \cite{dalvi2011centralized}\\
CloneCloud & Utilizes computation offloading through elastic execution between mobile devices and cloud. & \cite{chun2011clonecloud}\\

LWAPP (RFC 5412) & Proposes lightweight access point protocol (LWAPP) for centralized cloud-based WLAN management. & \cite{calhoun2010lightweight}\\

CloudMAC & Enables APs to redirect MAC frames only. Processing of MAC data is done via cloud computing infrastructure. & \cite{dely2012cloudmac} \\

IEEE 802.11 on Cloud-based Radio over Fibre & Conducts a study on the feasibility of the architecture of IEEE 802.11 on cloud-based radio over fibre.  & \cite{zhang2014feasibility} \\

 \vspace{1mm}
 \textbf{\emph{\underline{Cognitive Wireless Networks}}}\\ 
 
Coexistence of 802.15.4 with IEEE 802.11 & Proposes distributed adaptation strategies to ensure coexistence of IEEE 802.11 WLAN and IEEE 802.15.4 wireless sensor networks (WSNs) in the ISM band. &\cite{pollin2006distributed} \\

Coexistence between IEEE 802.11b and IEEE 802.16a networks & Proposes algorithms---based on dynamic frequency selection (DFS), power control (PC) and time-agility (TA)---to allow IEEE 802.11b and IEEE 802.16a networks to coexist in the same unlicensed band. & \cite{jing2005reactive} \\

QoS-aware MAC for IEEE 802.11p & Proposes an efficient multichannel QoS cognitive MAC (MQOG) for cognitive vehicular networks. &\cite{ajaltouni2012efficient} \\

Integration of IEEE 802.11 and 3G & Proposes schemes for integrating IEEE 802.11 and 3G seamlessly while satisfying QoS guarantees and roaming agreements.
&\cite{buddhikot2003integration} \\

Managing TCP in DSA-based WLANs & Proposes a framework known as DSASync for improved end-to-end TCP performance in dynamic spectrum access (DSA) wireless networks. & \cite{kumar2010managing} \\

Handover between IEEE 802.11b and overlay networks & Proposes algorithms for intersystem handover between IEEE 802.11b and an overlay network while satisfying the QoS parameters of minimum data rate, maximum data block delay, and maximum BER. & \cite{garmonov2008qos} \\

 \\

\bottomrule

\end{tabular}
\label{tab:example}
\end{table*}

\subsection{Software-defined Wireless Networks (SWNs)}

With increasing deployment and diversification of wireless technology, managing wireless networks has become very challenging. Software-defined networking (SDN) is a promising architecture that can be used for conveniently operating, controlling, and managing wireless networks. The defining characteristic of SDN is generally understood to serve as the separation of the control and data planes. The presence of programmable controllers to adjust the operating parameters enables us to call these networks `software defined'.

\begin{figure}[!ht]
\centering
{
 \includegraphics[width=.35\textwidth]{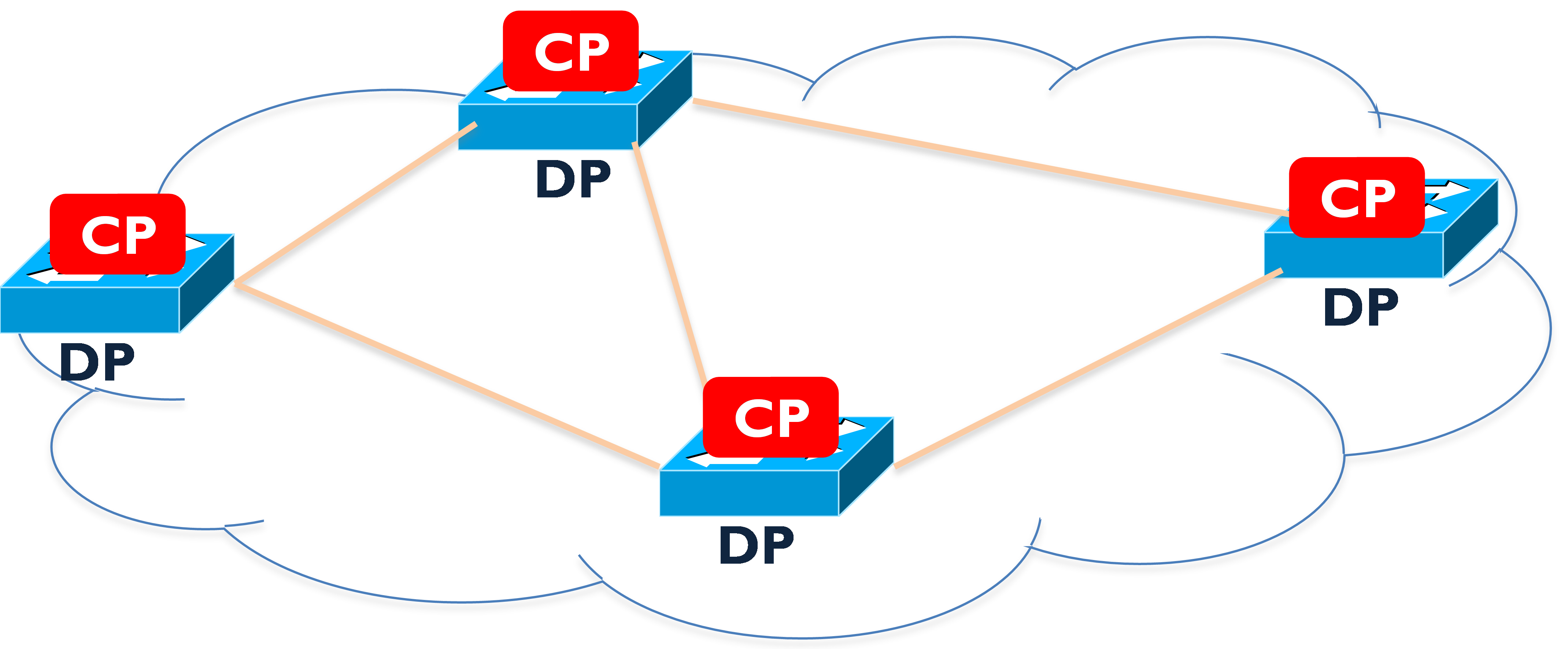}
  \label{fig:PreSDN}
 }
{
  \includegraphics[width=.45\textwidth]{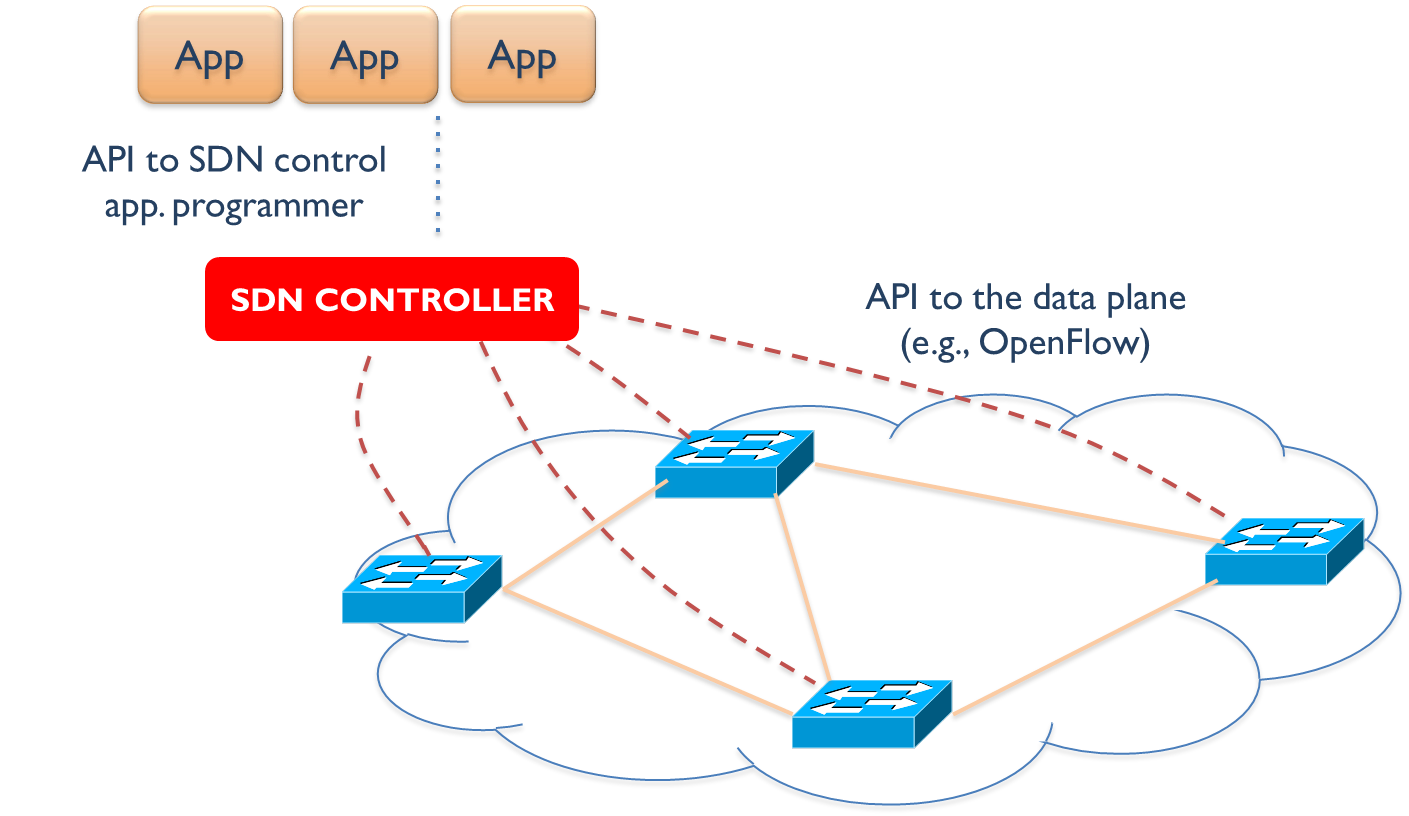}
  \label{fig:SDN}
}
\caption{Comparison of a traditional network and a SDN network architecture \cite{qadir2013building}.}
\label{fig:SDNoverall}
\end{figure}

Traditionally, networking devices, such as firewalls, routers, etc., require vendor-specific software for programming their operating parameters. This programming can be done manually by a network administrator through the command line interface (CLI). This limits the margin of innovation that can be incorporated into the modern networks such as the world wide web or the WLANs. SDN changes this notion of network programming by extracting the control intelligence from the data plane and managing all the data plane devices at centralized controller(s) \cite{mendoncca2013survey}. Figure \ref{fig:SDNoverall} shows a traditional network where the control and data planes are co-located in each networking device, and so the network control is decentralized. In comparison, SDN has a centralized architecture where the central SDN controller is controlling the multiple data planes; specifically, southbound APIs are used to communicate with data plane, and northbound APIs are used to communicate with SDN applications. SDN provides the flexibility of programming a network through the control plane. This can help in simplifying network management and operations. The rest of this subsection presents QoS efforts for IEEE 802.11-based SWNs.

\vspace{2mm}
\emph{\textbf{QoS efforts for IEEE 802.11-based SWNs}}: Traditionally, SDN has been focused on campus and data center networks; but increasingly, researchers are also focusing on wireless access networks. Since it can be difficult to manually configure various QoS knobs, OpenFlow and SDN architectures can be very useful in automating scalable control of network QoS based on high-level descriptions of application/service requirements. Kim et al. \cite{kim2010automated} proposed an OpenFlow-based network QoS control framework which includes per-flow rate limiters and dynamic priority assignment. 

SDN-enabled IEEE 802.11 networks provide the benefits of agile QoS provisioning. In an SDN network, bandwidth allocation, rate limiting, and traffic shaping can be efficiently implemented at the network level through automated QoS network APIs via the network controller. Real-time measurements can help to steer policies that can efficiently enforce the QoS mechanisms in real time \cite{{amanisdn}}. QoS policies can be deployed at the controller and the wastage of resources can be minimized. In a SDN-enabled network, several controllers in a single area can communicate with each other to exchange network information. This communication allows the users to connect with APs, regardless of their operators, thus improving the user's quality of experience (QoE). An AP that receives packets with a destination address of another network can forward the packets flexibly through rules defined at the controller \cite{{chaudet2013wireless}}. 

Ishimori et al. \cite{{ishimori2013control}} proposed a QoS solution called QoSFlow for OpenFlow-based SDN networks through the appropriate control of packet scheduling. OpenFlow is an archetypal SDN protocol used for implementing the architectural vision of separated control and data planes. OpenFlow implements a protocol used by the SDN controller to communicate with controlled devices. OpenFlow provides basic QoS primitives including support of only FIFO scheduling which may be insufficient for some applications like multimedia streaming. QoS support for OpenFlow is improving: queues are available in OpenFlow 1.0 which enables traffic shaping, while in the latest version, namely OpenFlow 1.3, rate limiting can be supported through meter tables. This paper proposed QoSFlow as a QoS development strategy that relies on multiple packet schedulers for OpenFlow supported networks to overcome the limitations of FIFO packet scheduling. QoSFlow can provide control for the following packet schedulers: hierarchical token bucket (HTB), random early detection (RED) and stochastic fair queueing (SFQ).

OpenQoS \cite{{egilmez2012koc}} is an OpenFlow controller designed for supporting multimedia flows with end-to-end QoS requirements. It enables QoS by placing multimedia traffic on QoS-guaranteed routes. OpenQoS presents a new dynamic QoS routing scheme that maintains the shortest path for the data delivery, which helps in minimizing packet loss and latency. The results show that the network turbulence has a minimal effect on video quality with QoS support. On the other hand, the videos without QoS support suffer significantly from quality degradation.

A dynamic framework for ensuring QoS in streaming videos at the control plane is presented in \cite{{egilmez2013optimization}}. The scheme works in the OpenFlow-based networks by optimizing forwarding decisions at the SDN controller. The controller acts as the brain of a network where the forwarding decisions are made. The routing choices are associated with the priority of the data flows. The resources are reserved at the controller, based on the type of delivery the controller can provide \cite{{sezer2013we}}. This reservation scheme does not affect the other types of flows due to the dynamic routing mechanism in the OpenFlow architecture.

Zhao et al. define a framework for a single SDN controller controlling all the APs via OpenFlow interface in \cite{{zhao2014leveraging}}. The proposed framework adds specific rules in various APs for packet scheduling without modifying the conventional DCF mechanism. Lee et al. in \cite{lee2014mesdn} extend the SDN controller network to mobile devices. It helps in achieving real-time detection of QoS demands in a network and can provide end-to-end QoS control.

In \cite{{kassler2012towards}}, the QoS measurements are taken at the service level and the network level. At the network level, the data flows from a source to a destination along the same path. This helps the media to be delivered in the best possible path and service configuration. In return, the overall QoS for the users is improved. SDNs are proposed to combine the configurations of network elements and end hosts. This enables network operators to define their own set of rules to control the traffic routing and QoS.

The QoS routing significantly affects the overall throughput of a network. The major portion of Internet traffic is composed of best-effort traffic. Therefore, an efficient QoS routing algorithm must take into account the existence of best-effort traffic and its impact on the overall performance. The OpenFlow controllers can generate flow tables to manage the QoS within a network using different routing protocols \cite{{civanlar2010qos}}. The controller performs additional functions of QoS contract management and QoS route management. The controller also polls switches to report on the congestion level in the network.

\subsection{Cloud-based Wireless Networks (CbWNs)}

\begin{figure}[t]
\begin{center}
\includegraphics[width=.45\textwidth]{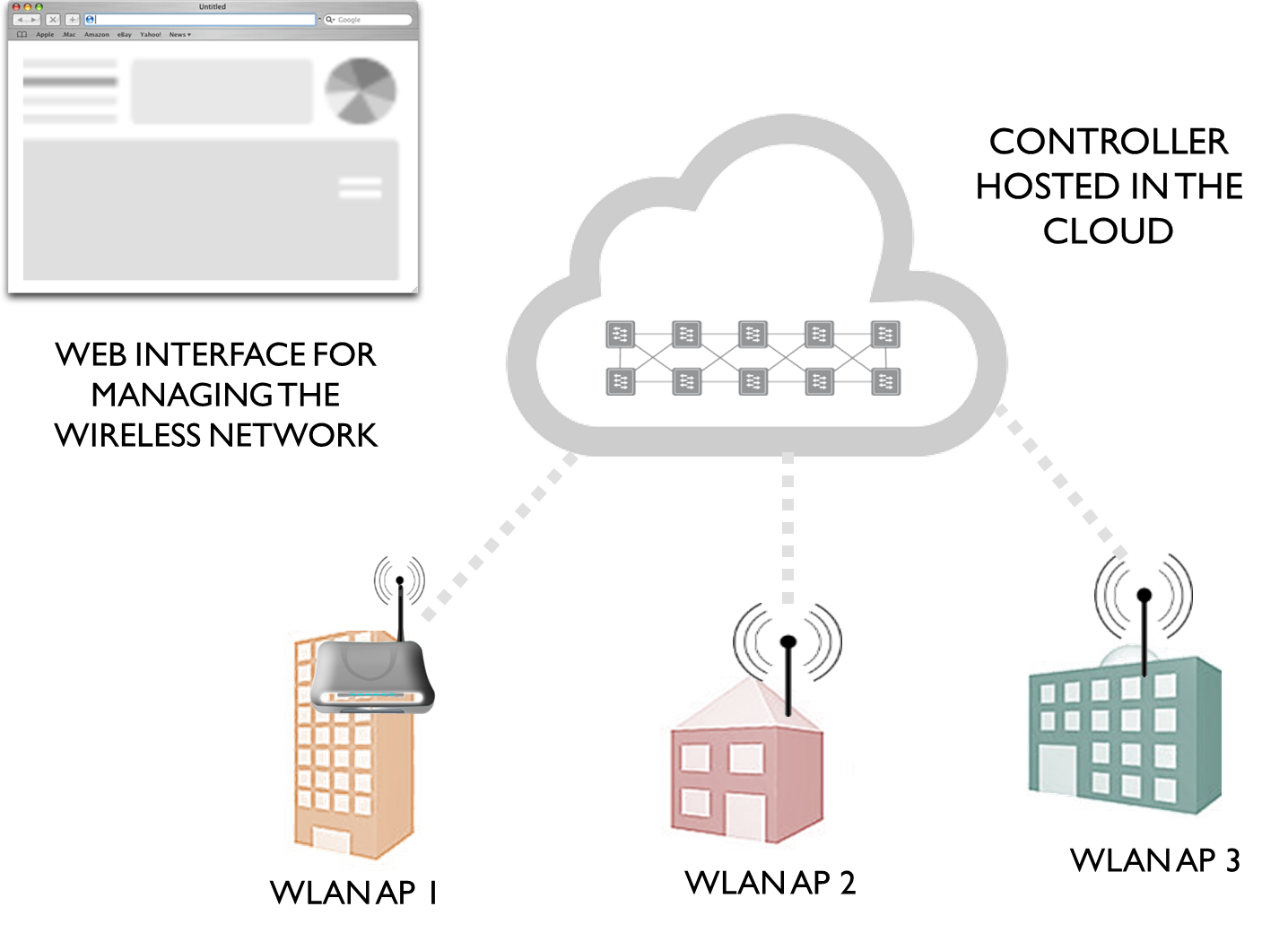}
\caption{Cloud-based Wireless Networks (CbWNs) \cite{qadir2013building}.} 
\label{fig:CbWN}
\end{center}
\end{figure}

The main idea of cloud computing is to offer computing services (e.g., networks, storage, applications)---provisioned through APIs via the web from a shared pool of resources---in a virtualized data center in utility computing style \cite{Armbrust2010}. The capability to program a network through the cloud/SDN architectures allow revolutionary dynamism in service provisioning, network management and control. The trend of CbWNs is to extend the cloud computing concept to wireless networks \cite{qadir2013building}. An illustration of the CbWN architecture is shown in Figure \ref{fig:CbWN}. Some salient advantages of CbWNs are described next.

\begin{itemize}
\vspace{1mm}
\item \emph{Centralized Management:} A major application of CbWNs is \emph{centralized remote management} of wireless networks. The control and provisioning of wireless access points (CAPWAP) protocol---defined by IETF in RFC 5415 \cite{capwap2009}---is a control and management (C\&M) protocol that aims at migrating functionalities from the hardware AP equipment to an external controller potentially managed via the cloud. There exists significant interest in the research community in proposing efficient approaches for central management of Wi-Fi networks \cite{dalvi2011centralized}. Various industrial solutions, such as Meraki Networks \cite{meraki} from Cisco, Aruba Networks \cite{aruba}, and AeroHive, have also been proposed to perform cloud-based management of WLANs.

\vspace{1mm}
\item \emph{Zero-Touch Auto-Configuration:}
The centralized management paradigm of CbWNs can allow plug and play \emph{zero-touch auto-configuration} of wireless APs allowing the APs to function without any manual configuration by the network administrator. The centralized management of wireless networks also allow cloud-based performance management and the use of \emph{advanced data analytics} for optimization performance including \emph{real-time reconfiguration} of wireless parameters.
\end{itemize}

\vspace{2mm}
\emph{\textbf{QoS efforts for IEEE 802.11-based CbWNs}}: Most of the devices used for cloud computing applications are portable and are connected through IEEE 802.11 WLAN. The WiFi access network may not be able to cope with the need of providing services to QoS-assured cloud multimedia applications. In particular, when the devices are in motion, QoS-aware handover should take into account the traffic load and currently accessible bandwidth at each location at particular AP with high precision. Consequently, a precise modeling of the genuine surroundings of IEEE 802.11 WLAN is important for proficient QoS-aware cloud service provisioning. Tursunova et al. proposed a pragmatic IEEE 802.11e EDCA model for QoS-aware differentiated multimedia cloud service provisioning in WLAN networks \cite{tursunova2012realistic}. 

Most of the previous work concentrated on the analysis of EDCA in saturated and non-saturated states. In \cite{engelstad2005non} \cite{malone2007mac} \cite{giustiniano2010measuring} \cite{wu2006ieee} \cite{ kosek2011simple}, the authors examine only clean channel surroundings, with an assumption that the frame error is caused by packet collision only, while in real surroundings, frame errors may take place due to channel noise. In \cite{tursunova2012realistic}, the authors considered the time-varying frame error probability of independent stations. They strengthened the mathematical model of IEEE 802.11e \cite{kosek2011simple} by using the calculated packet error probability, which may be caused by frame collisions and channel noise. Experimental outcomes show that the suggested model gives more precise assessment, compared to existing analytical models.\\

WLANs deployed by large firms or universities can compose of hundreds or even thousands of APs. Similarly the size of the operating system of an AP also increases with time due to the inclusion of software packages in each release. Therefore, it is getting harder for the network administrators to configure each AP individually. Reducing complexity of networking appliances and uncovering data flow management tasks via standardized interfaces and high-level programming primitives are some of the main concepts of SDN. In \cite{{sharkh2013resource}}, the authors present a resource allocation mechanism based on the cloud environments, as well as an energy-aware model for the data centers. 

To obtain similar advantages in WLANs, Dely et al. introduced CloudMAC \cite{dely2012cloudmac}, which is a novel management architecture in which access points redirect MAC frames only. The rest of the functionalities, like the processing of MAC data or management frames, is executed in typical servers that are operated in data centers and can be allocated via cloud computing infrastructure. OpenFlow is used to organize the flow and transmission characteristics of MAC frames.

In another work, Chun et al. proposed CloneCloud which allows unmodified mobile applications running in an application-level virtual memory (VM) to seamlessly offload part of its execution from the mobile devices onto device clones operating in the cloud \cite{chun2011clonecloud}. The ability to offload computation can be exploited in a QoS framework to meet stringent deadlines. 

\subsection{Cognitive Wireless Networks (CWN)}
\label{sec:CWN}

\begin{figure}[t]
\begin{center}
\includegraphics[width=.4\textwidth]{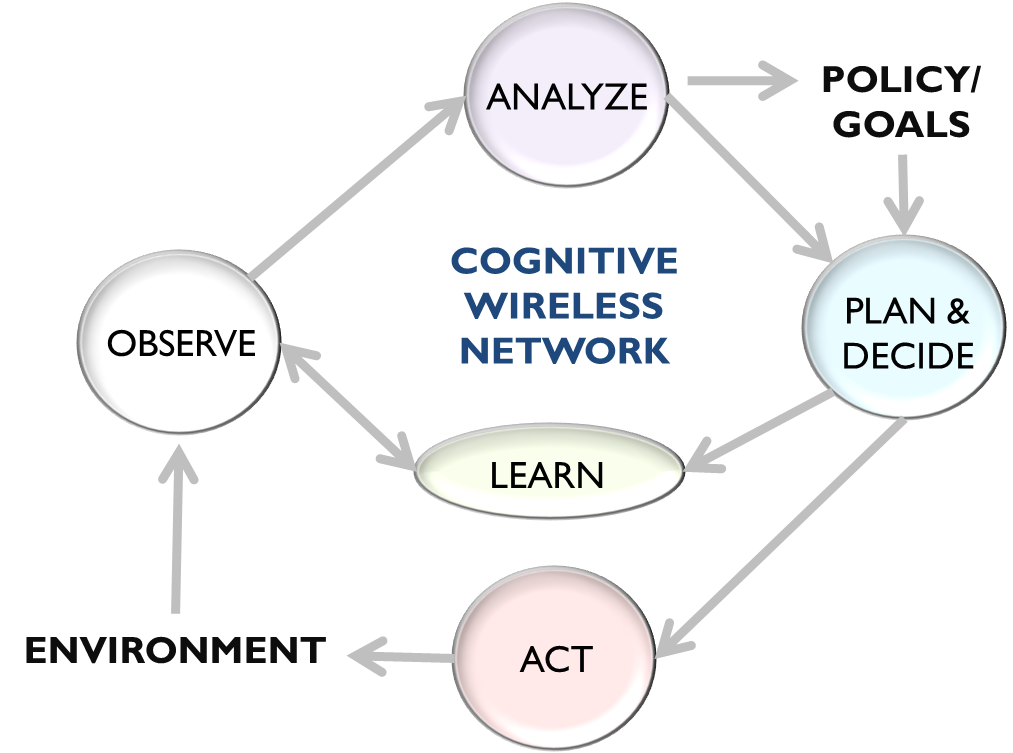}
\caption{Cognitive wireless networks (CWN) include an embedded cognitive engine which can observe network conditions, orient itself with the context, learn from experience, and decide to act \cite{qadir2013building}.}
\label{fig:ooda}
\end{center}
\end{figure}

Cognitive wireless networks (CWNs) are next-generation wireless networks---that demonstrate network-wide intelligent behavior---in which network nodes are incorporated with cognitive engines (see Figure \ref{fig:ooda}) consist of substantial artificial intelligence (AI) approaches in the form of machine learning, knowledge reasoning, optimization, and natural language processing \cite{akyildiz2006next} \cite{haykin2005cognitive}. Such networks are composed of network nodes equipped with cognitive radios (CR), which display device-level intelligent behavior. 

Along with network-level reconfiguration capabilities afforded by technologies such as SDN which can be used to realize programmable data plane and programmable control plane, future wireless programmable networking will also implement some variant of a ``\emph{knowledge plane}'' \cite{clark2003knowledge}. Traditionally, a network of nodes equipped with CRs is called a cognitive radio network (CRN) with the dominant application of CR technology being dynamic spectrum access (DSA), which can resolve the `artificial spectrum scarcity' problem resulting from the classic command-and-control licensing approach \cite{fette2009cognitive} adopted in various countries around the world. Since CRNs inherently embody AI techniques with wireless communications, it seems natural to explore using CRs to provide mechanisms for implementing the knowledge plane of future programmable wireless devices.

While the bulk of CRN work has focused on enabling device-level intelligent behavior, the concept of CWNs, initially proposed in \cite{thomas2007cognitive}, generalizes CRNs and emphasizes \emph{network-level} intelligence and self-aware behavior. While DSA is the most popularly cited application of CRNs, developing network-level intelligence in CRNs enables numerous other applications---including the ability to reprogram itself optimally according to network conditions.

In previous CRN research, it has been observed that PHY and MAC layers offer many ``\emph{knobs}'' that can be tweaked to optimize performance which can be measured through some ``\emph{meters}''. In \cite{fette2009cognitive}, many examples of knobs and meters at the PHY and MAC layers have been provided. Since CRNs operate in dynamic, often unknown, conditions, configuring the knobs optimally is not a trivial problem. Various AI-based techniques have been proposed in the literature to assist CRNs in their quest of performing autonomous optimal adaptations in such settings. Apart from AI techniques, CRN also borrows techniques and tools from various other fields such as game theory, control theory, optimization theory, metaheuristics, etc. \cite{haykin2005cognitive}.

Game theory has been used in various work to model, analyze, and develop QoS solutions for CRNs. Berlemann et al. \cite{berlemann2005radio} have proposed the use of radio resource sharing games to enable distributed QoS solutions in unlicensed bands shared by multiple users. Attar et al. proposed a game-theoretic resource allocation framework that guarantees QoS in a DSA environment (in which the primary network is assumed to be OFDM-based cellular network). The QoS is defined by the minimum rate available to the primary network and the target BER. 

Optimization theory has also been used in the literature to address the problem of QoS-constrained dynamic spectrum access. For example, Xing et al. \cite{xing2007dynamic} considered QoS differentiation for various unlicensed users while incorporating interference temperature constraints assuming a spectrum underlay access (alternatively, known as a shared-use model). In addition, various cross-layered solutions have been developed for ensuring QoS in CRNs. For example, Su and Zhang \cite{su2008cross} proposed a distributed cross-layered solution incorporating spectrum sensing at the PHY layer and packet scheduling at the MAC layer for QoS provisioning in CRNs. 

There has also been work in using cognitive technologies to facilitate QoS-aware coexistence among multiple 802.11 WLANs, between 802.11 and 802.16 networks \cite{berlemann2006coexistence} \cite{jing2005reactive}, and between 802.11 WLANs and overlay networks \cite{garmonov2008qos}.

\section{Open Research Issues and Future Work}
\label{sec:futurework}

QoS enhancement schemes in modern wireless-based networks still need further attention. This section highlights some of the important issues.

\subsection{Convergence of Different Technologies}

Wireless technologies are proliferating at a breakneck pace, and in such a dynamic ecosystem, technologies that facilitate multi-technology convergence is becoming increasingly important. In the future, IEEE 802.11-based networks will increasingly coexist with other wireless technologies such as 4G/5G, WiMAX, RFID, Internet of Things (IoT), wireless sensor networks, etc. As an example, 5G---expected to materialize by 2020---will be highly integrative and will tie Wi-Fi with other wireless mobile standards such as 3G and LTE \cite{andrews2014will}. IEEE 802.11-based wireless networks is facing stiff competition from other technologies that coexist in the unlicensed spectrum space, such as the IEEE 802.16-based wireless metropolitan area networks. Management of handover, spectrum sharing, coexistence, and interworking of diverse technologies, become important for ensuring QoS. There will be a lot of interest in QoS-aware spectrum sharing and coexistence between IEEE 802.11-based wireless networks and other technologies. This entails work at various layers including the potential use of multi-path TCP at the transport layer for improving QoS by exploiting multi-homing with a diverse range of wireless networks, such as Wi-Fi and 3G. More research needs to be done to ensure QoS in such environments where heterogeneous technologies exist.

\subsection{Context-Awareness and QoE}

To improve the QoS and QoE, it is imperative for researchers to seamlessly incorporate user preferences, and awareness of context, which can be based on \textit{identity}, \textit{location}, \textit{time}, or \textit{activity}, into IEEE-based networks. Since the end user's traffic varies with time, traffic behavior should be analyzed to predict the future traffic patterns and subsequently to adopt appropriate strategies. This helps in fulfilling the requirement of end devices with higher efficiency. Also, since the wireless networks are mobile, so predicting the future locations of nodes helps in data forwarding, and thus reducing the overall delay. If the future location of a node can be predicted from its mobility pattern and its speed, this helps in successful delivery of packets. Capturing the mobility patterns and its behavior ensure enhanced QoS.

\subsection{Challenges due to Virtualization}

Virtualization has transformed both operational efficiency and the economics of the computing industry, and more recently, the data center environment. With the growing role of virtualization in networking, it is highly likely that IEEE 802.11-based networks will increasingly utilize virtualization technology in the future \cite{qadir2013building}. In particular, the combination of cloud computing and network virtualization (including network functions virtualization) allow programmability that leads to unprecedented flexibility in rapidly creating, deploying, and managing novel services in virtualized settings as per the demands of users. This can create a new service-oriented architecture for wireless networking where heterogeneous wireless access technologies including IEEE 802.11 may coexist and converge as \emph{extended cloud infrastructure} \cite{wen2013wirelessbook}. The QoS challenges associated with virtualization/cloud management of IEEE 802.11 WLANs need to be investigated thoroughly to resolve potential issues.

\subsection{Cognitive Wireless Networking and QoS-Awareness}

Although some work has been done in the design of QoS-aware cognitive routing \cite{qadir2013artificial} \cite{how2011routing} and cognitive transport layer protocols \cite{kumar2010managing} \cite{chowdhury2009tp} for CWNs, more work needs to be done to realize the considerable promise of using AI- and machine learning-based techniques for developing IEEE 802.11-based wireless networks that can provide high QoS and QoE. 

\section{Conclusion}
\label{sec:conclusion}

It is anticipated that the bulk of access technologies in the near future will be wireless. With emerging standards such as 5G, and the convergence of the telecom and Internet industries on IP-based technologies, the ability to provide high QoS has become paramount. In this paper, we have surveyed various QoS enhancement techniques proposed for IEEE 802.11-based wireless networks with our discussion encompassing both classical techniques as well as proposals for new and emerging architectures such as SDN and cloud- networks. We have classified these techniques using different criteria. We have discussed QoS solutions that have been proposed for various TCP/IP layers along with a discussion on cross-layered protocols. We have also highlighted open research challenges and directions for future work.

%

\bibliographystyle{IEEEtran}
\bibliography{QoS}

\begin{thebibliography}{100}
\providecommand{\url}[1]{#1}
\csname url@samestyle\endcsname
\providecommand{\newblock}{\relax}
\providecommand{\bibinfo}[2]{#2}
\providecommand{\BIBentrySTDinterwordspacing}{\spaceskip=0pt\relax}
\providecommand{\BIBentryALTinterwordstretchfactor}{4}
\providecommand{\BIBentryALTinterwordspacing}{\spaceskip=\fontdimen2\font plus
\BIBentryALTinterwordstretchfactor\fontdimen3\font minus
  \fontdimen4\font\relax}
\providecommand{\BIBforeignlanguage}[2]{{%
\expandafter\ifx\csname l@#1\endcsname\relax
\typeout{** WARNING: IEEEtran.bst: No hyphenation pattern has been}%
\typeout{** loaded for the language `#1'. Using the pattern for}%
\typeout{** the default language instead.}%
\else
\language=\csname l@#1\endcsname
\fi
#2}}
\providecommand{\BIBdecl}{\relax}
\BIBdecl

\bibitem{baghaei2004review}
N.~Baghaei and R.~Hunt, ``Review of quality of service performance in wireless
  {LAN}s and {3G} multimedia application services,'' \emph{Computer
  Communications}, vol.~27, no.~17, pp. 1684--1692, 2004.

\bibitem{gast2005802}
M.~Gast, \emph{802.11 wireless networks: the definitive guide}.\hskip 1em plus
  0.5em minus 0.4em\relax " O'Reilly Media, Inc.", 2005.

\bibitem{wang2001internet}
Z.~Wang, \emph{Internet QoS: architectures and mechanisms for quality of
  service}.\hskip 1em plus 0.5em minus 0.4em\relax Morgan Kaufmann, 2001.

\bibitem{chen2004qos}
D.~Chen and P.~K. Varshney, ``{{QoS}} support in wireless sensor networks: A
  survey.'' in \emph{International Conference on Wireless Networks}, vol. 233,
  2004, pp. 1--7.

\bibitem{mogre2007qos}
P.~S. Mogre, M.~Hollick, and R.~Steinmetz, ``{{QoS}} in wireless mesh networks:
  challenges, pitfalls, and roadmap to its realization,'' in
  \emph{International workshop on Network and Operating Systems Support for
  Digital Audio \& Video (NOSSDAV)}, 2007.

\bibitem{zhu2004survey}
H.~Zhu, M.~Li, I.~Chlamtac, and B.~Prabhakaran, ``A survey of quality of
  service in {IEEE} 802.11 networks,'' \emph{Wireless Communications, IEEE},
  vol.~11, no.~4, pp. 6--14, 2004.

\bibitem{lindgren2001evaluation}
A.~Lindgren, A.~Almquist, and O.~Schel{\'e}n, ``Evaluation of quality of
  service schemes for {IEEE} 802.11 wireless lans,'' in \emph{Local Computer
  Networks, 2001. Proceedings. LCN 2001. 26th Annual IEEE Conference on}.\hskip
  1em plus 0.5em minus 0.4em\relax IEEE, 2001, pp. 348--351.

\bibitem{ni2004survey}
Q.~Ni, L.~Romdhani, and T.~Turletti, ``A survey of {{QoS}} enhancements for
  {IEEE} 802.11 wireless {LAN},'' \emph{Wireless Communications and Mobile
  Computing}, vol.~4, no.~5, pp. 547--566, 2004.

\bibitem{aboul2009wireless}
O.~Aboul-Magd, \emph{Wireless Local Area Networks Quality of Service: An
  Engineering Perspective}.\hskip 1em plus 0.5em minus 0.4em\relax IEEE, 2009.

\bibitem{Wang2001}
Z.~Wang, \emph{Internet {QoS}: architectures and mechanisms for quality of
  service}.\hskip 1em plus 0.5em minus 0.4em\relax Morgan Kaufmann, 2001.

\bibitem{jha2002engineering}
S.~Jha and M.~Hassan, \emph{Engineering Internet {QoS}}.\hskip 1em plus 0.5em
  minus 0.4em\relax Artech House, 2002.

\bibitem{setton2005cross}
E.~Setton, T.~Yoo, X.~Zhu, A.~Goldsmith, and B.~Girod, ``Cross-layer design of
  ad hoc networks for real-time video streaming,'' \emph{Wireless
  Communications, IEEE}, vol.~12, no.~4, pp. 59--65, 2005.

\bibitem{aurrecoechea1998survey}
C.~Aurrecoechea, A.~T. Campbell, and L.~Hauw, ``A survey of {{QoS}}
  architectures,'' \emph{Multimedia systems}, vol.~6, no.~3, pp. 138--151,
  1998.

\bibitem{mahadevan1999quality}
I.~Mahadevan and K.~M. Sivalingam, ``Quality of service architectures for
  wireless networks: Intserv and diffserv models,'' in \emph{Parallel
  Architectures, Algorithms, and Networks, 1999.(I-SPAN'99) Proceedings. Fourth
  InternationalSymposium on}.\hskip 1em plus 0.5em minus 0.4em\relax IEEE,
  1999, pp. 420--425.

\bibitem{zhang1993rsvp}
L.~Zhang, S.~Deering, D.~Estrin, S.~Shenker, and D.~Zappala, ``{RSVP}: A new
  resource reservation protocol,'' \emph{Network, IEEE}, vol.~7, no.~5, pp.
  8--18, 1993.

\bibitem{clark1992supporting}
D.~D. Clark, S.~Shenker, and L.~Zhang, \emph{Supporting real-time applications
  in an integrated services packet network: Architecture and mechanism}.\hskip
  1em plus 0.5em minus 0.4em\relax ACM, 1992, vol.~22, no.~4.

\bibitem{shenkerspecification}
S.~Shenker, C.~Partridge, and R.~Guerin, ``Specification of guaranteed quality
  of service, {RFC} 2212, september 1997,'' Internet RFC 2212, Tech. Rep.

\bibitem{braden1994integrated}
R.~Braden, D.~Clark, S.~Shenker \emph{et~al.}, ``Integrated services in the
  internet architecture: an overview. {RFC} 1633.'' 1994.

\bibitem{blake1998architecture}
S.~Blake, D.~Black, M.~Carlson, E.~Davies, Z.~Wang, and W.~Weiss, ``An
  architecture for differentiated services, {RFC} 2475.'' \emph{RFC 2475},
  1998.

\bibitem{jacobson1999expedited}
V.~Jacobson, K.~Nichols, K.~Poduri \emph{et~al.}, ``An expedited forwarding
  {PHB}. {RFC} 2598.'' \emph{RFC 2598}, 1999.

\bibitem{nichols1999two}
K.~Nichols and V.~Jacobson, ``A two-bit differentiated services architecture
  for the internet,'' \emph{{RFC} 2638}, 1999.

\bibitem{chaouchi2004adaptive}
H.~Chaouchi and A.~Munaretto, ``Adaptive {{QoS}} management for {IEEE} 802.11
  future wireless {ISPs},'' \emph{Wireless Networks}, vol.~10, no.~4, pp.
  413--421, 2004.

\bibitem{garcia2003quality}
J.~A. Garc{\i}a-Mac{\i}as, F.~Rousseau, G.~Berger-Sabbatel, L.~Toumi, and
  A.~Duda, ``Quality of service and mobility for the wireless {I}nternet,''
  \emph{Wireless Networks}, vol.~9, no.~4, pp. 341--352, 2003.

\bibitem{christin2003qos}
N.~Christin and J.~Liebeherr, ``A {{QoS}} architecture for quantitative service
  differentiation,'' \emph{Communications Magazine, IEEE}, vol.~41, no.~6, pp.
  38--45, 2003.

\bibitem{aad2001differentiation}
I.~Aad and C.~Castelluccia, ``Differentiation mechanisms for {IEEE} 802.11,''
  in \emph{INFOCOM 2001. Twentieth Annual Joint Conference of the {IEEE}
  Computer and Communications Societies. Proceedings. IEEE}, vol.~1.\hskip 1em
  plus 0.5em minus 0.4em\relax IEEE, 2001, pp. 209--218.

\bibitem{ksentini2004adaptive}
A.~Ksentini, M.~Naimi, A.~Nafaa \emph{et~al.}, ``Adaptive service
  differentiation for {{QoS}} provisioning in {IEEE} 802.11 wireless ad hoc
  networks,'' in \emph{Proceedings of the 1st ACM international workshop on
  Performance evaluation of wireless ad hoc, sensor, and ubiquitous
  networks}.\hskip 1em plus 0.5em minus 0.4em\relax ACM, 2004, pp. 39--45.

\bibitem{hanzo2007survey}
L.~Hanzo and R.~Tafazolli, ``A survey of {{QoS}} routing solutions for mobile
  ad hoc networks,'' \emph{IEEE Communications Surveys \& Tutorials}, vol.~9,
  no. 2 2nd, pp. 50--70, 2007.

\bibitem{chalmers1999survey}
D.~Chalmers and M.~Sloman, ``A survey of quality of service in mobile computing
  environments,'' \emph{Communications Surveys \& Tutorials, IEEE}, vol.~2,
  no.~2, pp. 2--10, 1999.

\bibitem{tanenbaum2003computer}
A.~S. Tanenbaum, ``Computer networks, 4-th edition,'' 2003.

\bibitem{knightly1999admission}
E.~W. Knightly and N.~B. Shroff, ``Admission control for statistical {{QoS}} :
  {Theory} and practice,'' \emph{Network, IEEE}, vol.~13, no.~2, pp. 20--29,
  1999.

\bibitem{perros1996call}
H.~G. Perros and K.~M. Elsayed, ``Call admission control schemes: a review,''
  \emph{IEEE Communications Magazine}, vol.~34, no.~11, pp. 82--91, 1996.

\bibitem{gao2005admission}
D.~Gao, J.~Cai, and K.~N. Ngan, ``Admission control in {IEEE 802.11} e wireless
  lans,'' \emph{Network, IEEE}, vol.~19, no.~4, pp. 6--13, 2005.

\bibitem{hou2009theory}
I.-H. Hou, V.~Borkar, and P.~Kumar, ``A theory of {QoS} for wireless,'' in
  \emph{INFOCOM 2009, IEEE}, April 2009, pp. 486--494.

\bibitem{vicisano1998tcp}
L.~Vicisano, J.~Crowcroft, and L.~Rizzo, ``{TCP}-like congestion control for
  layered multicast data transfer,'' in \emph{INFOCOM'98. Seventeenth Annual
  Joint Conference of the {IEEE} Computer and Communications Societies.
  Proceedings. IEEE}, vol.~3.\hskip 1em plus 0.5em minus 0.4em\relax IEEE,
  1998, pp. 996--1003.

\bibitem{xiao1999internet}
X.~Xiao and L.~M. Ni, ``Internet qos: a big picture,'' \emph{Network, IEEE},
  vol.~13, no.~2, pp. 8--18, 1999.

\bibitem{keshav1997engineering}
S.~Keshav, ``An engineering approach to computer networking: {ATM} networks,
  the internet, and the telephone network,'' \emph{Reading MA}, vol. 11997,
  1997.

\bibitem{peterson2007computer}
L.~L. Peterson and B.~S. Davie, \emph{Computer networks: a systems
  approach}.\hskip 1em plus 0.5em minus 0.4em\relax Elsevier, 2007.

\bibitem{hiertz2010ieee}
G.~R. Hiertz, D.~Denteneer, L.~Stibor, Y.~Zang, X.~P. Costa, and B.~Walke,
  ``The {IEEE} 802.11 universe,'' \emph{Communications Magazine, IEEE},
  vol.~48, no.~1, pp. 62--70, 2010.

\bibitem{dujovne2010taxonomy}
D.~Dujovne, T.~Turletti, and F.~Filali, ``A taxonomy of {IEEE} 802.11 wireless
  parameters and open source measurement tools,'' \emph{Communications Surveys
  \& Tutorials, IEEE}, vol.~12, no.~2, pp. 249--262, 2010.

\bibitem{ng2005experimental}
A.~C. Ng, D.~Malone, and D.~J. Leith, ``Experimental evaluation of {TCP}
  performance and fairness in an 802.11 e test-bed,'' in \emph{Proceedings of
  the 2005 ACM SIGCOMM workshop on Experimental approaches to wireless network
  design and analysis}.\hskip 1em plus 0.5em minus 0.4em\relax ACM, 2005, pp.
  17--22.

\bibitem{ni2005performance}
Q.~Ni, ``Performance analysis and enhancements for {IEEE} 802.11e wireless
  networks,'' \emph{Network, IEEE}, vol.~19, no.~4, pp. 21--27, 2005.

\bibitem{wu2001streaming}
D.~Wu, Y.~T. Hou, W.~Zhu, Y.-Q. Zhang, and J.~M. Peha, ``Streaming video over
  the internet: approaches and directions,'' \emph{Circuits and Systems for
  Video Technology, {IEEE} Transactions on}, vol.~11, no.~3, pp. 282--300,
  2001.

\bibitem{Mangold2002}
S.~Mangold, S.~Choi, P.~May, O.~Klein, G.~Hiertz, and L.~Stibor, ``{IEEE}
  802.11e wireless {LAN} for quality of service,'' in \emph{Proc. European
  Wireless}, vol.~2, 2002, pp. 32--39.

\bibitem{lamaire1996wireless}
R.~O. LaMaire, A.~Krishna, P.~Bhagwat, and J.~Panian, ``Wireless {LAN}s and
  mobile networking: standards and future directions,'' \emph{Communications
  Magazine, IEEE}, vol.~34, no.~8, pp. 86--94, 1996.

\bibitem{raniwala2005architecture}
A.~Raniwala and T.-c. Chiueh, ``Architecture and algorithms for an {IEEE}
  802.11-based multi-channel wireless mesh network,'' in \emph{INFOCOM 2005.
  24th Annual Joint Conference of the {IEEE} Computer and Communications
  Societies. Proceedings IEEE}, vol.~3.\hskip 1em plus 0.5em minus 0.4em\relax
  IEEE, 2005, pp. 2223--2234.

\bibitem{zhou2006crowded}
G.~Zhou, J.~A. Stankovic, and S.~H. Son, ``Crowded spectrum in wireless sensor
  networks,'' \emph{IEEE EmNets}, vol.~6, 2006.

\bibitem{mohapatra2003qos}
P.~Mohapatra, J.~Li, and C.~Gui, ``{{QoS}} in mobile ad hoc networks,''
  \emph{IEEE Wireless Communications}, vol.~10, no.~3, pp. 44--53, 2003.

\bibitem{guo2002class}
Y.~Guo and H.~Chaskar, ``Class-based quality of service over air interfaces in
  {4G} mobile networks,'' \emph{Communications Magazine, IEEE}, vol.~40, no.~3,
  pp. 132--137, 2002.

\bibitem{charfi2013phy}
E.~Charfi, L.~Chaari, and L.~Kamoun, ``{PHY/MAC} enhancements and {{QoS}}
  mechanisms for very high throughput {WLAN}s: A survey,'' \emph{Communications
  Surveys \& Tutorials, IEEE}, vol.~15, no.~4, pp. 1714--1735, 2013.

\bibitem{li2007real}
Y.~Li, C.~S. Chen, Y.-Q. Song, Z.~Wang \emph{et~al.}, ``Real-time {{QoS}}
  support in wireless sensor networks: a survey,'' in \emph{7th IFAC
  International Conference on Fieldbuses \& Networks in Industrial \& Embedded
  Systems-FeT'2007}, 2007.

\bibitem{zhao2002performance}
J.~Zhao, Z.~Guo, Q.~Zhang, and W.~Zhu, ``Performance study of {MAC} for service
  differentiation in {IEEE} 802.11,'' in \emph{Global Telecommunications
  Conference, 2002. GLOBECOM'02. IEEE}, vol.~1.\hskip 1em plus 0.5em minus
  0.4em\relax IEEE, 2002, pp. 778--782.

\bibitem{li2005}
M.~Li and B.~Prabhakaran, ``{MAC} layer admission control and priority
  re-allocation for handling {{QoS}} guarantees in non-cooperative wireless
  lans,'' \emph{Mobile networks and applications}, vol.~10, no.~6, pp.
  947--959, 2005.

\bibitem{achary2012enhanced}
R.~Achary, P.~R. Chellaih, V.~Vaityanathan, and S.~Nagarajan, ``Enhanced
  {{QoS}} by service differentiation in {MAC}-layer for {WLAN}.''
  \emph{International Journal of Computer Applications}, vol.~55, 2012.

\bibitem{sundareswaran2007improving}
N.~Sundareswaran, G.~F. Riley, K.~Boyd, and A.~Nainani, ``Improving quality of
  service in {MAC} 802.11 layer,'' in \emph{Modeling, Analysis, and Simulation
  of Computer and Telecommunication Systems, 2007. MASCOTS'07. 15th
  International Symposium on}.\hskip 1em plus 0.5em minus 0.4em\relax IEEE,
  2007, pp. 38--45.

\bibitem{ni2004qos}
Q.~Ni and T.~Turletti, ``{{QoS}} support for {IEEE} 802.11 wireless {LAN},''
  \emph{INRIA, Sophia Antipolis, France}, 2004.

\bibitem{lindgren2003quality}
A.~Lindgren, A.~Almquist, and O.~Schel{\'e}n, ``Quality of service schemes for
  {IEEE} 802.11 wireless {LAN}s: an evaluation,'' \emph{Mobile Networks and
  Applications}, vol.~8, no.~3, pp. 223--235, 2003.

\bibitem{drabu1999survey}
Y.~Drabu, ``A survey of {{QoS}} techniques in 802.11,'' \emph{Internet'Online},
  pp. 0001--03, 1999.

\bibitem{aad2000introducing}
I.~Aad and C.~Castelluccia, ``Introducing service differentiation into {IEEE}
  802.11,'' in \emph{Computers and Communications, 2000. Proceedings. ISCC
  2000. Fifth IEEE Symposium on}.\hskip 1em plus 0.5em minus 0.4em\relax IEEE,
  2000, pp. 438--443.

\bibitem{sharma2013quality}
V.~Sharma, J.~Malhotra, and H.~Singh, ``Quality of service (qos) evaluation of
  {IEEE 802.11} {WLAN} using different phy-layer standards,''
  \emph{Optik-International Journal for Light and Electron Optics}, vol. 124,
  no.~4, pp. 357--360, 2013.

\bibitem{wang2011mobility}
Z.~Wang, T.~Jiang, L.~Zhang, and Y.~Liu, ``Mobility and {QoS} oriented 802.11p
  {MAC} scheme for vehicle-to-infrastructure communications,'' in
  \emph{Communications and Networking in China (CHINACOM), 2011 6th
  International ICST Conference on}.\hskip 1em plus 0.5em minus 0.4em\relax
  IEEE, 2011, pp. 669--674.

\bibitem{chen2011high}
Y.-S. Chen, M.-Y. Chuang, F.-C. Tseng, and C.-H. Ke, ``High performance
  distributed coordination function with {{QoS}} support in {IEEE 802.11e}
  networks,'' in \emph{Australasian Telecommunication Networks and Applications
  Conference (ATNAC), 2011}.\hskip 1em plus 0.5em minus 0.4em\relax IEEE, 2011,
  pp. 1--6.

\bibitem{kowalski2013hybrid}
J.~Kowalski, ``Hybrid coordination in an {IEEE 802.11} network,'' Jun.~25 2013,
  uS Patent 8,472,416.

\bibitem{gargusing}
P.~Garg, R.~Doshi, R.~Greene, M.~Baker, M.~Malek, and X.~Cheng, ``Using {IEEE}
  802.11e {MAC} for {{QoS}} over wireless,'' in \emph{Performance, Computing,
  and Communications Conference, 2003. Conference Proceedings of the 2003
  {IEEE} International}.\hskip 1em plus 0.5em minus 0.4em\relax IEEE, 2003, pp.
  537--542.

\bibitem{romdhani2003adaptive}
L.~Romdhani, Q.~Ni, and T.~Turletti, ``Adaptive {EDCF}: enhanced service
  differentiation for {IEEE} 802.11 wireless ad-hoc networks,'' in
  \emph{Wireless Communications and Networking, 2003. WCNC 2003. 2003 IEEE},
  vol.~2.\hskip 1em plus 0.5em minus 0.4em\relax IEEE, 2003, pp. 1373--1378.

\bibitem{qashi2011evaluating}
R.~Qashi, M.~Bogdan, and K.~Hanssgen, ``Evaluating the {{QoS}} of {WLANs} for
  the {IEEE 802.11} {EDCF} in real-time applications,'' in \emph{Communications
  and Information Technology (ICCIT), 2011 International Conference on}.\hskip
  1em plus 0.5em minus 0.4em\relax IEEE, 2011, pp. 32--35.

\bibitem{yu2013resource}
X.~Yu, P.~Navaratnam, and K.~Moessner, ``Resource reservation schemes for
  {IEEE} 802.11-based wireless networks: A survey,'' \emph{Communications
  Surveys \& Tutorials, IEEE}, vol.~15, no.~3, pp. 1042--1061, 2013.

\bibitem{zhang2009qos}
D.-y. Zhang, J.~Jiang, A.~Anani, and H.-b. Li, ``{QoS}-guaranteed packet
  scheduling in wireless networks,'' \emph{The Journal of China Universities of
  Posts and Telecommunications}, vol.~16, no.~2, pp. 63--67, 2009.

\bibitem{lu1999fair}
S.~Lu, V.~Bharghavan, and R.~Srikant, ``Fair scheduling in wireless packet
  networks,'' \emph{IEEE/ACM Transactions on Networking (TON)}, vol.~7, no.~4,
  pp. 473--489, 1999.

\bibitem{tsao2000extending}
S.-L. Tsao, ``Extending earliest-due-date scheduling algorithms for wireless
  networks with location-dependent errors,'' in \emph{Vehicular Technology
  Conference, 2000. IEEE-VTS Fall VTC 2000. 52nd}, vol.~1.\hskip 1em plus 0.5em
  minus 0.4em\relax IEEE, 2000, pp. 223--228.

\bibitem{grilo2003scheduling}
A.~Grilo, M.~Macedo, and M.~Nunes, ``A scheduling algorithm for {{QoS}} support
  in {IEEE}802. 11 networks,'' \emph{Wireless Communications, IEEE}, vol.~10,
  no.~3, pp. 36--43, 2003.

\bibitem{lim2004qos}
L.~Lim, R.~Malik, P.~Tan, C.~Apichaichalermwongse, K.~Ando, and Y.~Harada, ``A
  {{QoS}} scheduler for {IEEE} 802.11e {WLANs},'' in \emph{Consumer
  Communications and Networking Conference, 2004. CCNC 2004. First IEEE}.\hskip
  1em plus 0.5em minus 0.4em\relax IEEE, 2004, pp. 199--204.

\bibitem{xiao2004ieee}
Y.~Xiao, ``{IEEE} 802.11e: {QoS} provisioning at the {MAC} layer,''
  \emph{Wireless Communications, IEEE}, vol.~11, no.~3, pp. 72--79, 2004.

\bibitem{ansel2004efficient}
P.~Ansel, Q.~Ni, and T.~Turletti, ``An efficient scheduling scheme for {IEEE}
  802.11e,'' in \emph{Proc. Modeling and Optimization in Mobile, Ad Hoc and
  Wireless Networks}, 2004, pp. 24--26.

\bibitem{georges2004formal}
J.-P. Georges, T.~Divoux, and E.~Rondeau, ``A formal method to guarantee a
  deterministic behaviour of switched ethernet networks for time-critical
  applications,'' in \emph{Computer Aided Control Systems Design, 2004 {IEEE}
  International Symposium on}.\hskip 1em plus 0.5em minus 0.4em\relax IEEE,
  2004, pp. 255--260.

\bibitem{jasperneite2002deterministic}
J.~Jasperneite, P.~Neumann, M.~Theis, and K.~Watson, ``Deterministic real-time
  communication with switched {Ethernet},'' in \emph{Proceedings of the 4th
  {IEEE} International Workshop on Factory Communication Systems}.\hskip 1em
  plus 0.5em minus 0.4em\relax Citeseer, 2002, pp. 11--18.

\bibitem{georges2005strict}
J.-P. Georges, T.~Divoux, and E.~Rondeau, ``Strict priority versus weighted
  fair queueing in switched ethernet networks for time critical applications,''
  in \emph{Parallel and Distributed Processing Symposium, 2005. Proceedings.
  19th {IEEE} International}.\hskip 1em plus 0.5em minus 0.4em\relax IEEE,
  2005, pp. 141--141.

\bibitem{jiang2002probabilistic}
Y.~Jiang, C.-K. Tham, and C.-C. Ko, ``A probabilistic priority scheduling
  discipline for multi-service networks,'' \emph{Computer Communications},
  vol.~25, no.~13, pp. 1243--1254, 2002.

\bibitem{parekh1993generalized}
A.~K. Parekh and R.~G. Gallager, ``A generalized processor sharing approach to
  flow control in integrated services networks: the single-node case,''
  \emph{IEEE/ACM Transactions on Networking (ToN)}, vol.~1, no.~3, pp.
  344--357, 1993.

\bibitem{chen2005interleaved}
Y.-M. Chen, H.~Chung, E.~Lee, and Z.~Tong, ``Interleaved weighted fair queuing
  mechanism and system,'' Dec.~13 2005, uS Patent 6,975,638.

\bibitem{banchs2002distributed}
A.~Banchs and X.~Perez, ``Distributed weighted fair queuing in 802.11 wireless
  lan,'' in \emph{Communications, 2002. ICC 2002. {IEEE} International
  Conference on}, vol.~5.\hskip 1em plus 0.5em minus 0.4em\relax IEEE, 2002,
  pp. 3121--3127.

\bibitem{kwon1998scheduling}
T.-G. Kwon, S.-H. Lee, and J.-K. Rho, ``Scheduling algorithm for real-time
  burst traffic using dynamic weighted round robin,'' in \emph{Circuits and
  Systems, 1998. ISCAS'98. Proceedings of the 1998 IEEE International Symposium
  on}, vol.~6.\hskip 1em plus 0.5em minus 0.4em\relax IEEE, 1998, pp. 506--509.

\bibitem{kwak2002modified}
J.-Y. Kwak, J.-S. Nam, and D.-H. Kim, ``A modified dynamic weighted round robin
  cell scheduling algorithm,'' \emph{ETRI journal}, vol.~24, no.~5, pp.
  360--372, 2002.

\bibitem{elsayed2006channel}
K.~M. Elsayed and A.~K. Khattab, ``Channel-aware earliest deadline due fair
  scheduling for wireless multimedia networks,'' \emph{Wireless Personal
  Communications}, vol.~38, no.~2, pp. 233--252, 2006.

\bibitem{morris2008automatic}
K.~J. Morris, D.~J. Hudson, and A.~Goyal, ``Automatic adaptive network traffic
  prioritization and shaping,'' Dec.~2 2008, uS Patent 7,460,476.

\bibitem{zhang2007qos}
D.~Zhang and D.~Ionescu, ``{QoS} performance analysis in deployment of
  {D}iffserv-aware {MPLS} traffic engineering,'' in \emph{Software Engineering,
  Artificial Intelligence, Networking, and Parallel/Distributed Computing,
  2007. SNPD 2007. Eighth ACIS International Conference on}, vol.~3.\hskip 1em
  plus 0.5em minus 0.4em\relax IEEE, 2007, pp. 963--967.

\bibitem{mangold2003analysis}
S.~Mangold, S.~Choi, G.~R. Hiertz, O.~Klein, and B.~Walke, ``Analysis of {IEEE}
  802.11e for {{QoS}} support in wireless {LANs},'' \emph{Wireless
  Communications, IEEE}, vol.~10, no.~6, pp. 40--50, 2003.

\bibitem{choi2003ieee}
S.~Choi, J.~Del~Prado, N.~Sai~Shankar, and S.~Mangold, ``Ieee 802.11e
  contention-based channel access (edcf) performance evaluation,'' in
  \emph{Communications, 2003. ICC'03. {IEEE} International Conference on},
  vol.~2.\hskip 1em plus 0.5em minus 0.4em\relax IEEE, 2003, pp. 1151--1156.

\bibitem{yang2002priority}
X.~Yang and N.~H. Vaidya, ``Priority scheduling in wireless ad hoc networks,''
  in \emph{Proceedings of the 3rd ACM international symposium on Mobile ad hoc
  networking \& computing}.\hskip 1em plus 0.5em minus 0.4em\relax ACM, 2002,
  pp. 71--79.

\bibitem{krithikaquality}
P.~Krithika and M.~Pushpavalli, ``Quality of service optimization in {IEEE}
  802.11e networks using enhanced distributed channel access techniques,''
  \emph{International Journal of Computer Networks and Wireless Communications
  (IJCNWC)}, 2012.

\bibitem{villalonprovisioning}
J.~Villal{\'o}n, F.~Mic{\'o}, P.~Cuenca, and L.~Orozco-Barbosa, ``Provisioning
  {QoS} support for multimedia traffic in {IEEE} 802.11 {WLAN}s: A performance
  evaluation.''

\bibitem{khanoptimization}
M.~Khan, T.~A. Khan, and M.~Beg, ``Optimization of wireless network {MAC} layer
  parameters,'' \emph{International Journal of Innovative Technology and
  Exploring Engineering (IJITEE)}, 2013.

\bibitem{rashid2008controlled}
M.~M. Rashid, E.~Hossain, and V.~K. Bhargava, ``Controlled channel access
  scheduling for guaranteed {{QoS}} in 802.11 e-based {WLAN}s,'' \emph{Wireless
  Communications, {IEEE} Transactions on}, vol.~7, no.~4, pp. 1287--1297, 2008.

\bibitem{reddy2006quality}
T.~B. Reddy, I.~Karthigeyan, B.~Manoj, and C.~Murthy, ``Quality of service
  provisioning in ad hoc wireless networks: a survey of issues and solutions,''
  \emph{Ad Hoc Networks}, vol.~4, no.~1, pp. 83--124, 2006.

\bibitem{gu2004sequential}
D.~Gu, Y.~Yuan, and J.~Zhang, ``Sequential coordinated channel access in
  wireless networks,'' Jul.~10 2004, uS Patent App. 10/888,398.

\bibitem{rashid2007hcca}
M.~M. Rashid, E.~Hossain, and V.~K. Bhargava, ``{HCCA} scheduler design for
  guaranteed {{QoS}} in {IEEE} 802.11e based {WLANs},'' in \emph{Wireless
  Communications and Networking Conference, 2007. WCNC 2007. IEEE}.\hskip 1em
  plus 0.5em minus 0.4em\relax IEEE, 2007, pp. 1538--1543.

\bibitem{cervello2006collision}
G.~Cervello and S.~Choi, ``Collision avoidance in {IEEE 802.11} contention free
  period (cfp) with overlapping basic service sets ({BSSs}),'' May~30 2006, uS
  Patent 7,054,329.

\bibitem{yeh2002support}
J.-Y. Yeh and C.~Chen, ``Support of multimedia services with the {IEEE} 802.11
  {MAC} protocol,'' in \emph{Communications, 2002. ICC 2002. {IEEE}
  International Conference on}, vol.~1.\hskip 1em plus 0.5em minus 0.4em\relax
  IEEE, 2002, pp. 600--604.

\bibitem{perez2010ieee}
X.~P{\'e}rez-Costa and D.~Camps-Mur, ``{IEEE} 802.11e {{QoS}} and power saving
  features overview and analysis of combined performance [accepted from open
  call],'' \emph{Wireless Communications, IEEE}, vol.~17, no.~4, pp. 88--96,
  2010.

\bibitem{tinnirello2005efficiency}
I.~Tinnirello and S.~Choi, ``Efficiency analysis of burst transmissions with
  block {ACK} in contention-based 802.11e {WLAN}s,'' in \emph{Communications,
  2005. ICC 2005. 2005 {IEEE} International Conference on}, vol.~5.\hskip 1em
  plus 0.5em minus 0.4em\relax IEEE, 2005, pp. 3455--3460.

\bibitem{politis2011exploiting}
A.~Politis, I.~Mavridis, and A.~Manitsaris, ``Exploiting multimedia frame
  semantics and {MAC}-layer enhancements for {{QoS}} provisioning in {IEEE}
  802.11e congested networks,'' \emph{International Journal On Advances in
  Networks and Services}, vol.~4, no. 1 and 2, pp. 176--185, 2011.

\bibitem{mangold2002ieee}
S.~Mangold, S.~Choi, P.~May, O.~Klein, G.~Hiertz, and L.~Stibor, ``{IEEE}
  802.11e wireless {LAN} for quality of service,'' in \emph{Proc. European
  Wireless}, vol.~2, 2002, pp. 32--39.

\bibitem{xiao2004local}
Y.~Xiao and H.~Li, ``Local data control and admission control for {QoS} support
  in wireless ad hoc networks,'' \emph{Vehicular Technology, {IEEE}
  Transactions on}, vol.~53, no.~5, pp. 1558--1572, 2004.

\bibitem{andreadistechniques}
A.~Andreadis and R.~Zambon, ``Techniques for preserving {{QoS}} performance in
  contention-based {IEEE} 802.11e networks,'' \emph{Intech Open, Advanced
  Wireless LAN}, 2012.

\bibitem{hanzo2009admission}
I.~Hanzo and R.~Tafazolli, ``Admission control schemes for 802.11-based
  multi-hop mobile ad hoc networks: a survey,'' \emph{Communications Surveys \&
  Tutorials, IEEE}, vol.~11, no.~4, pp. 78--108, 2009.

\bibitem{brewer2010comparison}
O.~T. Brewer and A.~Ayyagari, ``Comparison and analysis of measurement and
  parameter based admission control methods for quality of service (qos)
  provisioning,'' in \emph{Military Communications Conference, 2010-Milcom
  2010}.\hskip 1em plus 0.5em minus 0.4em\relax IEEE, 2010, pp. 184--188.

\bibitem{nor2006admission}
S.~Nor, A.~Mohd, and C.~Cheow, ``An admission control method for {IEEE}
  802.11e,'' \emph{Network Theory and Applications}, pp. 105--122, 2006.

\bibitem{wu2010design}
H.-T. Wu, M.-H. Yang, and K.-W. Ke, ``The design of {QoS} provisioning
  mechanisms for wireless networks,'' in \emph{Pervasive Computing and
  Communications Workshops (PERCOM Workshops), 2010 8th {IEEE} International
  Conference on}.\hskip 1em plus 0.5em minus 0.4em\relax IEEE, 2010, pp.
  756--759.

\bibitem{chen2006supporting}
X.~Chen, H.~Zhai, X.~Tian, and Y.~Fang, ``Supporting {{QoS}} in {IEEE} 802.11e
  wireless {LAN}s,'' \emph{Wireless Communications, {IEEE} Transactions on},
  vol.~5, no.~8, pp. 2217--2227, 2006.

\bibitem{cano2007adaptive}
C.~Cano, B.~Bellalta, and M.~Oliver, ``Adaptive admission control mechanism for
  {IEEE} 802.11e {WLAN}s,'' in \emph{Personal, Indoor and Mobile Radio
  Communications, 2007. PIMRC 2007. IEEE 18th International Symposium
  on}.\hskip 1em plus 0.5em minus 0.4em\relax IEEE, 2007, pp. 1--5.

\bibitem{ksentini2007etxop}
A.~Ksentini, A.~Nafaa, A.~Gueroui, and M.~Naimi, ``{ETXOP}: A resource
  allocation protocol for {{QoS}} -sensitive services provisioning in 802.11
  networks,'' \emph{Performance Evaluation}, vol.~64, no.~5, pp. 419--443,
  2007.

\bibitem{bensaou2009measurement}
B.~Bensaou, Z.-N. Kong, and D.~H. Tsang, ``A measurement-assisted, model-based
  admission control algorithm for {IEEE} 802.11e,'' \emph{Journal of
  Interconnection Networks}, vol.~10, no.~04, pp. 303--320, 2009.

\bibitem{sivakumar1999cedar}
R.~Sivakumar, P.~Sinha, and V.~Bharghavan, ``{CEDAR}: a core-extraction
  distributed ad hoc routing algorithm,'' \emph{Selected Areas in
  Communications, {IEEE} Journal on}, vol.~17, no.~8, pp. 1454--1465, 1999.

\bibitem{yin2006traffic}
S.~Yin, Y.~Xiong, Q.~Zhang, and X.~Lin, ``Traffic-aware routing for real-time
  communications in wireless multi-hop networks,'' \emph{Wireless
  Communications and Mobile Computing}, vol.~6, no.~6, pp. 825--843, 2006.

\bibitem{matos2012quality}
R.~Matos, N.~Coutinho, C.~Marques, S.~Sargento, J.~Chakareski, and A.~Kassler,
  ``Quality of experience-based routing in multi-service wireless mesh
  networks,'' in \emph{Communications (ICC), 2012 IEEE International Conference
  on}.\hskip 1em plus 0.5em minus 0.4em\relax IEEE, 2012, pp. 7060--7065.

\bibitem{lin1999qos}
C.~R. Lin and J.-S. Liu, ``Qos routing in ad hoc wireless networks,''
  \emph{Selected Areas in Communications, IEEE Journal on}, vol.~17, no.~8, pp.
  1426--1438, 1999.

\bibitem{chen1999distributed}
S.~Chen and K.~Nahrstedt, ``Distributed quality-of-service routing in ad hoc
  networks,'' \emph{Selected Areas in Communications, IEEE Journal on},
  vol.~17, no.~8, pp. 1488--1505, 1999.

\bibitem{abdrabou2006position}
A.~Abdrabou and W.~Zhuang, ``A position-based {{QoS}} routing scheme for uwb
  mobile ad hoc networks,'' \emph{Selected Areas in Communications, IEEE
  Journal on}, vol.~24, no.~4, pp. 850--856, 2006.

\bibitem{bashandy2005generalized}
A.~R. Bashandy, E.~K. Chong, and A.~Ghafoor, ``Generalized quality-of-service
  routing with resource allocation,'' \emph{Selected Areas in Communications,
  IEEE Journal on}, vol.~23, no.~2, pp. 450--463, 2005.

\bibitem{wang2005application}
M.~Wang and G.-S. Kuo, ``An application-aware {{QoS}} routing scheme with
  improved stability for multimedia applications in mobile ad hoc networks,''
  in \emph{Vehicular Technology Conference, 2005. VTC-2005-Fall. 2005 IEEE
  62nd}, vol.~3.\hskip 1em plus 0.5em minus 0.4em\relax IEEE, 2005, pp.
  1901--1905.

\bibitem{chou2006low}
C.~T. Chou, A.~Misra, and J.~Qadir, ``Low-latency broadcast in multirate
  wireless mesh networks,'' \emph{Selected Areas in Communications, IEEE
  Journal on}, vol.~24, no.~11, pp. 2081--2091, 2006.

\bibitem{sheng2003routing}
M.~Sheng, J.~Li, and Y.~Shi, ``Routing protocol with {{QoS}} guarantees for
  ad-hoc network,'' \emph{Electronics Letters}, vol.~39, no.~1, pp. 143--145,
  2003.

\bibitem{rubin2003link}
I.~Rubin and Y.-C. Liu, ``Link stability models for {{QoS}} ad hoc routing
  algorithms,'' in \emph{Vehicular Technology Conference, 2003. VTC 2003-Fall.
  2003 IEEE 58th}, vol.~5.\hskip 1em plus 0.5em minus 0.4em\relax IEEE, 2003,
  pp. 3084--3088.

\bibitem{barolli2003qos}
L.~Barolli, A.~Koyama, and N.~Shiratori, ``A {{QoS}} routing method for ad-hoc
  networks based on genetic algorithm,'' in \emph{Database and Expert Systems
  Applications, 2003. Proceedings. 14th International Workshop on}.\hskip 1em
  plus 0.5em minus 0.4em\relax IEEE, 2003, pp. 175--179.

\bibitem{kim2004demand}
D.~Kim, C.-H. Min, and S.~Kim, ``On-demand {SIR} and bandwidth-guaranteed
  routing with transmit power assignment in ad hoc mobile networks,''
  \emph{Vehicular Technology, IEEE Transactions on}, vol.~53, no.~4, pp.
  1215--1223, 2004.

\bibitem{wisitpongphan2005qos}
N.~Wisitpongphan, G.~Ferrari, S.~Panichpapiboon, J.~Parikh, and O.~Tonguz,
  ``Qos provisioning using ber-based routing in ad hoc wireless networks,'' in
  \emph{Vehicular Technology Conference, 2005. VTC 2005-Spring. 2005 IEEE
  61st}, vol.~4.\hskip 1em plus 0.5em minus 0.4em\relax IEEE, 2005, pp.
  2483--2487.

\bibitem{toh2001maximum}
C.-K. Toh, ``Maximum battery life routing to support ubiquitous mobile
  computing in wireless ad hoc networks,'' \emph{Communications Magazine,
  IEEE}, vol.~39, no.~6, pp. 138--147, 2001.

\bibitem{liu2004courtesy}
W.~Liu, X.~Chen, Y.~Fang, and J.~M. Shea, ``Courtesy piggybacking: supporting
  differentiated services in multihop mobile ad hoc networks,'' \emph{Mobile
  Computing, {IEEE} Transactions on}, vol.~3, no.~4, pp. 380--393, 2004.

\bibitem{chen2005qos}
L.~Chen and W.~B. Heinzelman, ``{{QoS}} -aware routing based on bandwidth
  estimation for mobile ad hoc networks,'' \emph{Selected Areas in
  Communications, {IEEE} Journal on}, vol.~23, no.~3, pp. 561--572, 2005.

\bibitem{xue2003ad}
Q.~Xue and A.~Ganz, ``Ad hoc {QoS} on-demand routing {(AQOR)} in mobile ad hoc
  networks,'' \emph{Journal of parallel and distributed computing}, vol.~63,
  no.~2, pp. 154--165, 2003.

\bibitem{balakrishnan1997comparison}
H.~Balakrishnan, V.~N. Padmanabhan, S.~Seshan, and R.~H. Katz, ``A comparison
  of mechanisms for improving {TCP} performance over wireless links,''
  \emph{Networking, IEEE/ACM Transactions on}, vol.~5, no.~6, pp. 756--769,
  1997.

\bibitem{chen2002syndrome}
W.-P. Chen, Y.-C. Hsiao, J.~C. Hou, Y.~Ge, and M.~P. Fitz, ``Syndrome: a
  light-weight approach to improving {TCP} performance in mobile wireless
  networks,'' \emph{Wireless Communications and Mobile Computing}, vol.~2,
  no.~1, pp. 37--57, 2002.

\bibitem{jelassi2012quality}
S.~Jelassi, G.~Rubino, H.~Melvin, H.~Youssef, and G.~Pujolle, ``Quality of
  experience of {VoIP} service: A survey of assessment approaches and open
  issues,'' \emph{Communications Surveys \& Tutorials, IEEE}, vol.~14, no.~2,
  pp. 491--513, 2012.

\bibitem{bolot1998experience}
J.-C. Bolot and T.~Turletti, ``Experience with control mechanisms for packet
  video in the internet,'' \emph{ACM SIGCOMM Computer Communication Review},
  vol.~28, no.~1, pp. 4--15, 1998.

\bibitem{wu2000end}
D.~Wu, Y.~T. Hou, W.~Zhu, H.-J. Lee, T.~Chiang, Y.-Q. Zhang, and H.~J. Chao,
  ``On end-to-end architecture for transporting {MPEG-4} video over the
  internet,'' \emph{Circuits and Systems for Video Technology, {IEEE}
  Transactions on}, vol.~10, no.~6, pp. 923--941, 2000.

\bibitem{martins1996joint}
F.~C. Martins, W.~Ding, and E.~Feig, ``Joint control of spatial quantization
  and temporal sampling for very low bit rate video,'' in \emph{Acoustics,
  Speech, and Signal Processing, 1996. ICASSP-96. Conference Proceedings., 1996
  {IEEE} International Conference on}, vol.~4.\hskip 1em plus 0.5em minus
  0.4em\relax IEEE, 1996, pp. 2072--2075.

\bibitem{wiegand1996rate}
T.~Wiegand, M.~Lightstone, D.~Mukherjee, T.~G. Campbell, and S.~K. Mitra,
  ``Rate-distortion optimized mode selection for very low bit rate video coding
  and the emerging {H. 263} standard,'' \emph{Circuits and Systems for Video
  Technology, {IEEE} Transactions on}, vol.~6, no.~2, pp. 182--190, 1996.

\bibitem{ding1997joint}
W.~Ding, ``Joint encoder and channel rate control of {VBR} video over {ATM}
  networks,'' \emph{Circuits and Systems for Video Technology, {IEEE}
  Transactions on}, vol.~7, no.~2, pp. 266--278, 1997.

\bibitem{hsu1997joint}
C.-Y. Hsu, A.~Ortega, and A.~R. Reibman, ``Joint selection of source and
  channel rate for {VBR} video transmission under {ATM} policing constraints,''
  \emph{Selected Areas in Communications, {IEEE} Journal on}, vol.~15, no.~6,
  pp. 1016--1028, 1997.

\bibitem{jacobson1988congestion}
V.~Jacobson, ``Congestion avoidance and control,'' in \emph{ACM SIGCOMM
  Computer Communication Review}, vol.~18, no.~4.\hskip 1em plus 0.5em minus
  0.4em\relax ACM, 1988, pp. 314--329.

\bibitem{turletti1996videoconferencing}
T.~Turletti and C.~Huitema, ``Videoconferencing on the internet,''
  \emph{Networking, IEEE/ACM Transactions on}, vol.~4, no.~3, pp. 340--351,
  1996.

\bibitem{zhu2011intelligent}
R.~Zhu, ``Intelligent rate control for supporting real-time traffic in wlan
  mesh networks,'' \emph{Journal of Network and Computer Applications},
  vol.~34, no.~5, pp. 1449--1458, 2011.

\bibitem{nam2002experimental}
C.~H. Nam, S.~C. Liew, and C.~P. Fu, ``An experimental study of {ARQ} protocol
  in 802.11 b wireless {LAN},'' \emph{Proc. Wireless Personal Multimedia
  Communications (WPMC 2002)}, 2002.

\bibitem{crow1997ieee}
B.~P. Crow, I.~Widjaja, J.~G. Kim, and P.~T. Sakai, ``Ieee 802.11 wireless
  local area networks,'' \emph{Communications Magazine, IEEE}, vol.~35, no.~9,
  pp. 116--126, 1997.

\bibitem{liu1997error}
H.~Liu, H.~Ma, M.~El~Zarki, and S.~Gupta, ``Error control schemes for networks:
  An overview,'' \emph{Mobile Networks and Applications}, vol.~2, no.~2, pp.
  167--182, 1997.

\bibitem{aikawa1996forward}
S.~Aikawa, Y.~Motoyama, and M.~Umehira, ``Forward error correction schemes for
  wireless {ATM} systems,'' in \emph{Communications, 1996. ICC'96, Conference
  Record, Converging Technologies for Tomorrow's Applications. 1996 {IEEE}
  International Conference on}, vol.~1.\hskip 1em plus 0.5em minus 0.4em\relax
  IEEE, 1996, pp. 454--458.

\bibitem{choi2006ieee}
S.~Choi, Y.~Choi, and I.~Lee, ``{IEEE} 802.11 {MAC}-level {FEC} scheme with
  retransmission combining,'' \emph{Wireless Communications, {IEEE}
  Transactions on}, vol.~5, no.~1, pp. 203--211, 2006.

\bibitem{leith2005tcp}
D.~J. Leith, P.~Clifford, D.~Malone, and A.~Ng, ``{TCP} fairness in 802.11e
  {WLAN}s,'' \emph{IEEE Communications letters}, vol.~9, no.~11, pp. 964--966,
  2005.

\bibitem{van2006optimized}
M.~van~der Schaar, Y.~Andreopoulos, and Z.~Hu, ``{Optimized scalable video
  streaming over IEEE 802.11 a/e HCCA wireless networks under delay
  constraints},'' \emph{Mobile Computing, IEEE Transactions on}, vol.~5, no.~6,
  pp. 755--768, 2006.

\bibitem{chakareski2003rate}
J.~Chakareski and B.~Girod, ``Rate-distortion optimized packet scheduling and
  routing for media streaming with path diversity,'' in \emph{Data Compression
  Conference, 2003. Proceedings. DCC 2003}.\hskip 1em plus 0.5em minus
  0.4em\relax IEEE, 2003, pp. 203--212.

\bibitem{luo2006optimal}
H.~Luo, M.-L. Shyu, and S.-C. Chen, ``An optimal resource utilization scheme
  with end-to-end congestion control for continuous media stream
  transmission,'' \emph{Computer Networks}, vol.~50, no.~7, pp. 921--937, 2006.

\bibitem{luo2008video}
------, ``Video streaming over the internet with optimal bandwidth resource
  allocation,'' \emph{Multimedia Tools and Applications}, vol.~40, no.~1, pp.
  111--134, 2008.

\bibitem{li2004providing}
Q.~Li and M.~Van~der Schaar, ``Providing adaptive {{QoS}} to layered video over
  wireless local area networks through real-time retry limit adaptation,''
  \emph{Multimedia, IEEE Transactions on}, vol.~6, no.~2, pp. 278--290, 2004.

\bibitem{domingo2004interaction}
M.~C. Domingo and D.~Remondo, ``An interaction model between ad-hoc networks
  and fixed {IP} networks for {{QoS}} support,'' in \emph{Proceedings of the
  7th ACM international symposium on Modeling, analysis and simulation of
  wireless and mobile systems}.\hskip 1em plus 0.5em minus 0.4em\relax ACM,
  2004, pp. 188--194.

\bibitem{zhang2005qos}
B.~Zhang and H.~T. Mouftah, ``{QoS} routing for wireless ad hoc networks:
  problems, algorithms, and protocols,'' \emph{Communications Magazine, IEEE},
  vol.~43, no.~10, pp. 110--117, 2005.

\bibitem{van2004adaptive}
P.~Van~Beek, S.~Deshpande, H.~Pan, and I.~Sezan, ``Adaptive streaming of
  high-quality video over wireless lans,'' in \emph{Electronic Imaging
  2004}.\hskip 1em plus 0.5em minus 0.4em\relax International Society for
  Optics and Photonics, 2004, pp. 647--660.

\bibitem{chakareski2004application}
J.~Chakareski and P.~A. Chou, ``Application layer error-correction coding for
  rate-distortion optimized streaming to wireless clients,''
  \emph{Communications, IEEE Transactions on}, vol.~52, no.~10, pp. 1675--1687,
  2004.

\bibitem{majumda2002multicast}
A.~Majumda, D.~G. Sachs, I.~V. Kozintsev, K.~Ramchandran, and M.~M. Yeung,
  ``Multicast and unicast real-time video streaming over wireless lans,''
  \emph{Circuits and Systems for Video Technology, IEEE Transactions on},
  vol.~12, no.~6, pp. 524--534, 2002.

\bibitem{toumpis2003performance}
S.~Toumpis and A.~J. Goldsmith, ``Performance, optimization, and cross-layer
  design of media access protocols for wireless ad hoc networks,'' in
  \emph{Communications, 2003. ICC'03. IEEE International Conference on},
  vol.~3.\hskip 1em plus 0.5em minus 0.4em\relax IEEE, 2003, pp. 2234--2240.

\bibitem{indumathi2010adaptive}
G.~Indumathi and K.~Murugesan, ``An adaptive time slot allocation for
  statistical {{QoS}} guarantees in wireless networks using crosslayer
  approach.'' \emph{International Journal of Communication Networks \&
  Information Security}, vol.~2, no.~1, 2010.

\bibitem{liu2006cross}
Q.~Liu, X.~Wang, and G.~B. Giannakis, ``A cross-layer scheduling algorithm with
  {{QoS}} support in wireless networks,'' \emph{Vehicular Technology, {IEEE}
  Transactions on}, vol.~55, no.~3, pp. 839--847, 2006.

\bibitem{abd2006efficient}
S.~Abd El-atty, ``Efficient packet scheduling with pre-defined {{QoS}} using
  cross-layer technique in wireless networks,'' in \emph{Computers and
  Communications, 2006. ISCC'06. Proceedings. 11th IEEE Symposium on}.\hskip
  1em plus 0.5em minus 0.4em\relax IEEE, 2006, pp. 820--826.

\bibitem{sheng2011cooperative}
Z.~Sheng, K.~K. Leung, and Z.~Ding, ``Cooperative wireless networks: from radio
  to network protocol designs,'' \emph{Communications Magazine, IEEE}, vol.~49,
  no.~5, pp. 64--69, 2011.

\bibitem{girod1999feedback}
B.~Girod and N.~Farber, ``Feedback-based error control for mobile video
  transmission,'' \emph{Proceedings of the IEEE}, vol.~87, no.~10, pp.
  1707--1723, 1999.

\bibitem{farber1999analysis}
N.~Farber, K.~Stuhlmuller, and B.~Girod, ``Analysis of error propagation in
  hybrid video coding with application to error resilience,'' in \emph{Image
  Processing, 1999. ICIP 99. Proceedings. 1999 International Conference on},
  vol.~2.\hskip 1em plus 0.5em minus 0.4em\relax IEEE, 1999, pp. 550--554.

\bibitem{wu2007softmac}
H.~Wu, Y.~Liu, Q.~Zhang, and Z.-L. Zhang, ``Soft{MAC}: layer 2.5 collaborative
  {MAC} for multimedia support in multihop wireless networks,'' \emph{Mobile
  Computing, {IEEE} Transactions on}, vol.~6, no.~1, pp. 12--25, 2007.

\bibitem{zhang2008cross}
Q.~Zhang and Y.-Q. Zhang, ``Cross-layer design for {QoS} support in multihop
  wireless networks,'' \emph{Proceedings of the IEEE}, vol.~96, no.~1, pp.
  64--76, 2008.

\bibitem{wang2006omar}
J.~Wang, H.~Zhai, Y.~Fang, J.~M. Shea, and D.~Wu, ``{OMAR:} utilizing multiuser
  diversity in wireless ad hoc networks,'' \emph{Mobile Computing, {IEEE}
  Transactions on}, vol.~5, no.~12, pp. 1764--1779, 2006.

\bibitem{bucciol2004cross}
P.~Bucciol, G.~Davini, E.~Masala, E.~Filippi, and J.~C. De~Martin,
  ``Cross-layer perceptual {ARQ} for h.264 video streaming over 802.11 wireless
  networks,'' in \emph{Global Telecommunications Conference, 2004. GLOBECOM'04.
  IEEE}, vol.~5.\hskip 1em plus 0.5em minus 0.4em\relax IEEE, 2004, pp.
  3027--3031.

\bibitem{zander2013riding}
J.~Zander and P.~Mahonen, ``Riding the data tsunami in the cloud: myths and
  challenges in future wireless access,'' \emph{Communications Magazine, IEEE},
  vol.~51, no.~3, pp. 145--151, 2013.

\bibitem{zhang2005cross}
Q.~Zhang, F.~Yang, and W.~Zhu, ``Cross-layer {Q}o{S} {S}upport for {M}ultimedia
  {D}elivery over {W}ireless {I}nternet.'' \emph{EURASIP J. Adv. Sig. Proc.},
  vol. 2005, no.~2, pp. 207--219, 2005.

\bibitem{qu2006source}
Q.~Qu, Y.~Pei, J.~W. Modestino, and X.~Tian, ``Source-adaptation-based wireless
  video transport: a cross-layer approach,'' \emph{EURASIP Journal on Applied
  Signal Processing}, vol. 2006, pp. 260--260, 2006.

\bibitem{liu2005cross}
Q.~Liu, S.~Zhou, and G.~B. Giannakis, ``Cross-layer scheduling with prescribed
  qos guarantees in adaptive wireless networks,'' \emph{Selected Areas in
  Communications, IEEE Journal on}, vol.~23, no.~5, pp. 1056--1066, 2005.

\bibitem{agarwal2013optimal}
A.~Agarwal and A.~K. Jagannatham, ``Optimal adaptive modulation for {Q}o{S}
  constrained wireless networks with renewable energy sources,'' \emph{Wireless
  Communications Letters, IEEE}, vol.~2, no.~1, pp. 78--81, 2013.

\bibitem{xianyang2014design}
F.~Xianyang and W.~Feng, ``Design and {I}mplementation of
  {I}nterference-{A}ware {C}ooperative {Q}o{S} {R}outing for {M}ulti-hop
  {W}ireless {N}etwork,'' in \emph{Measuring Technology and Mechatronics
  Automation (ICMTMA), 2014 Sixth International Conference on}.\hskip 1em plus
  0.5em minus 0.4em\relax IEEE, 2014, pp. 211--217.

\bibitem{boutremans2003adaptive}
C.~Boutremans and J.-Y. Le~Boudec, ``Adaptive joint playout buffer and {FEC}
  adjustment for internet telephony,'' in \emph{INFOCOM 2003. Twenty-Second
  Annual Joint Conference of the {IEEE} Computer and Communications. {IEEE}
  Societies}, vol.~1.\hskip 1em plus 0.5em minus 0.4em\relax IEEE, 2003, pp.
  652--662.

\bibitem{tanigawa2011qos}
Y.~Tanigawa, J.-O. Kim, and H.~Tode, ``Qo{S}-{A}ware {R}etransmission with
  {N}etwork {C}oding based on {A}daptive {C}ooperation with {IEEE} 802.11e
  {EDCA},'' in \emph{Global Telecommunications Conference (GLOBECOM 2011), 2011
  IEEE}.\hskip 1em plus 0.5em minus 0.4em\relax IEEE, 2011, pp. 1--5.

\bibitem{zhu2007rate}
X.~Zhu, P.~Agrawal, J.~Pal~Singh, T.~Alpcan, and B.~Girod, ``Rate allocation
  for multi-user video streaming over heterogenous access networks,'' in
  \emph{Proceedings of the 15th international conference on Multimedia}.\hskip
  1em plus 0.5em minus 0.4em\relax ACM, 2007, pp. 37--46.

\bibitem{cruz2003optimal}
R.~L. Cruz and A.~V. Santhanam, ``Optimal routing, link scheduling and power
  control in multihop wireless networks,'' in \emph{INFOCOM 2003. Twenty-Second
  Annual Joint Conference of the {IEEE} Computer and Communications. {IEEE}
  Societies}, vol.~1.\hskip 1em plus 0.5em minus 0.4em\relax IEEE, 2003, pp.
  702--711.

\bibitem{moh2009link}
S.~Moh, ``Link quality aware route discovery for robust routing and high
  performance in mobile ad hoc networks,'' in \emph{High Performance Computing
  and Communications, 2009. HPCC'09. 11th IEEE International Conference
  on}.\hskip 1em plus 0.5em minus 0.4em\relax IEEE, 2009, pp. 281--288.

\bibitem{liu2003opportunistic}
Y.~Liu and E.~Knightly, ``Opportunistic fair scheduling over multiple wireless
  channels,'' in \emph{INFOCOM 2003. Twenty-Second Annual Joint Conference of
  the {IEEE} Computer and Communications. {IEEE} Societies}, vol.~2.\hskip 1em
  plus 0.5em minus 0.4em\relax IEEE, 2003, pp. 1106--1115.

\bibitem{wang2004opportunistic}
J.~Wang, H.~Zhai, Y.~Fang, and M.~C. Yuang, ``Opportunistic media access
  control and rate adaptation for wireless ad hoc networks,'' in
  \emph{Communications, 2004 IEEE International Conference on}, vol.~1.\hskip
  1em plus 0.5em minus 0.4em\relax IEEE, 2004, pp. 154--158.

\bibitem{kyasanur2006routing}
P.~Kyasanur and N.~H. Vaidya, ``Routing and link-layer protocols for
  multi-channel multi-interface ad hoc wireless networks,'' \emph{ACM SIGMOBILE
  Mobile Computing and Communications Review}, vol.~10, no.~1, pp. 31--43,
  2006.

\bibitem{lee2014mesdn}
J.~Lee, M.~Uddin, J.~Tourrilhes, S.~Sen, S.~Banerjee, M.~Arndt, K.-H. Kim, and
  T.~Nadeem, ``{meSDN}: Mobile extension of {SDN},'' \emph{Proceedings of the
  Fifth ACM workshop on Mobile cloud computing and services (MCS)}, 2014.

\bibitem{ishimori2013control}
A.~Ishimori, F.~Farias, E.~Cerqueira, and A.~Abel{\'e}m, ``Control of multiple
  packet schedulers for improving {{QoS}} on {OpenFlow/SDN} networking,'' in
  \emph{Software Defined Networks (EWSDN), 2013 Second European Workshop
  on}.\hskip 1em plus 0.5em minus 0.4em\relax IEEE, 2013, pp. 81--86.

\bibitem{egilmez2012koc}
H.~E. Egilmez, S.~T. Dane, K.~T. Bagci, and A.~M. Tekalp, ``Koc univ.,
  istanbul, turkey,'' in \emph{Signal \& Information Processing Association
  Annual Summit and Conference (APSIPA ASC), 2012 Asia-Pacific}.\hskip 1em plus
  0.5em minus 0.4em\relax IEEE, 2012, pp. 1--8.

\bibitem{kassler2012towards}
A.~Kassler, L.~Skorin-Kapov, O.~Dobrijevic, M.~Matijasevic, and P.~Dely,
  ``Towards {QoE}-driven multimedia service negotiation and path optimization
  with software defined networking,'' in \emph{Software, Telecommunications and
  Computer Networks (SoftCOM), 2012 20th International Conference on}.\hskip
  1em plus 0.5em minus 0.4em\relax IEEE, 2012, pp. 1--5.

\bibitem{sherwood2009flowvisor}
R.~Sherwood, G.~Gibb, K.-K. Yap, G.~Appenzeller, M.~Casado, N.~McKeown, and
  G.~Parulkar, ``Flowvisor: A network virtualization layer,'' \emph{OpenFlow
  Switch Consortium, Tech. Rep}, 2009.

\bibitem{egilmezdistributed}
H.~Egilmez and M.~Tekalp, ``Distributed {{QoS}} architectures for multimedia
  streaming over software defined networks,'' \emph{Multimedia, {IEEE}
  Transactions on}, vol.~16, no.~6, pp. 1597--1609, Oct 2014.

\bibitem{zhao2014leveraging}
D.~Zhao, M.~Zhu, and M.~Xu, ``Leveraging {SDN} and openflow to mitigate
  interference in enterprise wlan,'' \emph{Journal of Networks}, vol.~9, no.~6,
  pp. 1526--1533, 2014.

\bibitem{sonkoly2012qos}
B.~Sonkoly, A.~Guly{\'a}s, F.~N{\'e}meth, J.~Czentye, K.~Kurucz, B.~Novak, and
  G.~Vaszkun, ``On {{QoS}} support to {Ofelia} and {OpenFlow},'' in
  \emph{Software Defined Networking (EWSDN), 2012 European Workshop on}.\hskip
  1em plus 0.5em minus 0.4em\relax IEEE, 2012, pp. 109--113.

\bibitem{nam2013openqflow}
K.~Nam-Seok, H.~Hwanjo, P.~Jong-Dae, and P.~Hong-Shik, ``{OpenQFlow}: Scalable
  openflow with flow-based {{QoS}},'' \emph{IEICE Transactions on
  Communications}, vol.~96, no.~2, pp. 479--488, 2013.

\bibitem{tursunova2012realistic}
S.~Tursunova and Y.-T. Kim, ``Realistic {IEEE} 802.11e {EDCA} model for {{QoS}}
  -aware mobile cloud service provisioning,'' \emph{Consumer Electronics,
  {IEEE} Transactions on}, vol.~58, no.~1, pp. 60--68, 2012.

\bibitem{sharkh2013resource}
M.~A. Sharkh, M.~Jammal, A.~Shami, and A.~Ouda, ``Resource allocation in a
  network-based cloud computing environment: design challenges,''
  \emph{Communications Magazine, IEEE}, vol.~51, no.~11, pp. 46--52, 2013.

\bibitem{dalvi2011centralized}
A.~Dalvi, P.~Swamy, and B.~Meshram, ``Centralized management approach for
  {WLAN},'' in \emph{Computer Networks and Information Technologies}.\hskip 1em
  plus 0.5em minus 0.4em\relax Springer, 2011, pp. 578--580.

\bibitem{chun2011clonecloud}
B.-G. Chun, S.~Ihm, P.~Maniatis, M.~Naik, and A.~Patti, ``Clonecloud: elastic
  execution between mobile device and cloud,'' in \emph{Proceedings of the
  sixth conference on Computer systems}.\hskip 1em plus 0.5em minus 0.4em\relax
  ACM, 2011, pp. 301--314.

\bibitem{calhoun2010lightweight}
P.~Calhoun, ``{Lightweight access point protocol},''
  \url{http://tools.ietf.org/html/rfc5412}, 2010, accessed: 2013-09-12.

\bibitem{dely2012cloudmac}
P.~Dely, J.~Vestin, A.~Kassler, N.~Bayer, H.~Einsiedler, and C.~Peylo,
  ``{CloudMAC}: an {OpenFlow} based architecture for 802.11 {MAC} layer
  processing in the cloud,'' in \emph{Globecom Workshops (GC Wkshps), 2012
  IEEE}.\hskip 1em plus 0.5em minus 0.4em\relax IEEE, 2012, pp. 186--191.

\bibitem{zhang2014feasibility}
S.~Zhang and D.~R. Franklin, ``Feasibility study on the implementation of
  {IEEE} 802.11 on cloud-based radio over fibre architecture,'' in
  \emph{Communications (ICC), 2014 {IEEE} International Conference on}.\hskip
  1em plus 0.5em minus 0.4em\relax IEEE, 2014, pp. 2891--2896.

\bibitem{pollin2006distributed}
S.~Pollin, M.~Ergen, M.~Timmers, A.~Dejonghe, L.~Van~der Perre, F.~Catthoor,
  I.~Moerman, and A.~Bahai, ``Distributed cognitive coexistence of 802.15.4
  with 802.11,'' in \emph{Cognitive Radio Oriented Wireless Networks and
  Communications, 2006. 1st International Conference on}.\hskip 1em plus 0.5em
  minus 0.4em\relax IEEE, 2006, pp. 1--5.

\bibitem{jing2005reactive}
X.~Jing, S.-C. Mau, D.~Raychaudhuri, and R.~Matyas, ``Reactive cognitive radio
  algorithms for co-existence between {IEEE} 802.11b and 802.16a networks,'' in
  \emph{Global Telecommunications Conference, 2005. GLOBECOM'05. IEEE},
  vol.~5.\hskip 1em plus 0.5em minus 0.4em\relax IEEE, 2005, pp. 5--pp.

\bibitem{ajaltouni2012efficient}
H.~Ajaltouni, R.~W. Pazzi, and A.~Boukerche, ``An efficient {QoS MAC} for
  {IEEE} 802.11p over cognitive multichannel vehicular networks,'' in
  \emph{Communications (ICC), 2012 {IEEE} International Conference on}.\hskip
  1em plus 0.5em minus 0.4em\relax IEEE, 2012, pp. 413--417.

\bibitem{buddhikot2003integration}
M.~Buddhikot, G.~Chandranmenon, S.~Han, Y.-W. Lee, S.~Miller, and
  L.~Salgarelli, ``Integration of 802.11 and third-generation wireless data
  networks,'' in \emph{INFOCOM 2003. Twenty-Second Annual Joint Conference of
  the {IEEE} Computer and Communications. {IEEE} Societies}, vol.~1.\hskip 1em
  plus 0.5em minus 0.4em\relax IEEE, 2003, pp. 503--512.

\bibitem{kumar2010managing}
A.~Kumar and K.~G. Shin, ``Managing {TCP} connections in dynamic spectrum
  access based wireless {LANs},'' in \emph{Sensor Mesh and Ad Hoc
  Communications and Networks (SECON), 2010 7th Annual {IEEE} Communications
  Society Conference on}.\hskip 1em plus 0.5em minus 0.4em\relax IEEE, 2010,
  pp. 1--9.

\bibitem{garmonov2008qos}
A.~V. Garmonov, S.~H. Cheon, K.~L. Han, Y.~S. Park, A.~Savinkov, S.~Filin,
  S.~Moiseev, M.~Kondakov \emph{et~al.}, ``{{QoS}} -oriented intersystem
  handover between {IEEE} 802.11b and overlay networks,'' \emph{Vehicular
  Technology, {IEEE} Transactions on}, vol.~57, no.~2, pp. 1142--1154, 2008.

\bibitem{mendoncca2013survey}
B.~Nunes, M.~Mendonca, X.~Nguyen, K.~Obraczka, and T.~Turletti, ``A survey of
  software-defined networking: Past, present, and future of programmable
  networks,'' \emph{Communications Surveys Tutorials, IEEE}, vol.~PP, no.~99,
  pp. 1--18, 2014.

\bibitem{qadir2013building}
J.~Qadir, N.~Ahmed, and N.~Ahad, ``Building programmable wireless networks: An
  architectural survey,'' \emph{EURASIP Journal on Wireless Communications and
  Networking (EURASIP JWCN)}, 2014.

\bibitem{kim2010automated}
W.~Kim, P.~Sharma, J.~Lee, S.~Banerjee, J.~Tourrilhes, S.-J. Lee, and
  P.~Yalagandula, ``Automated and scalable {{QoS}} control for network
  convergence,'' \emph{Proc. INM/WREN}, vol.~10, pp. 1--1, 2010.

\bibitem{amanisdn}
M.~Amani, T.~Mahmoodi, M.~Tatipamula, and H.~Aghvami, ``{SDN}-based data
  offloading for {5G} mobile networks,'' \emph{ZTE Communications}, 2014.

\bibitem{chaudet2013wireless}
C.~Chaudet and Y.~Haddad, ``Wireless software defined networks: Challenges and
  opportunities,'' in \emph{Microwaves, Communications, Antennas and
  Electronics Systems (COMCAS), 2013 {IEEE} International Conference on}.\hskip
  1em plus 0.5em minus 0.4em\relax IEEE, 2013, pp. 1--5.

\bibitem{egilmez2013optimization}
H.~E. Egilmez, S.~Civanlar, and A.~M. Tekalp, ``An optimization framework for
  {{QoS}} -enabled adaptive video streaming over openflow networks,''
  \emph{Multimedia, {IEEE} Transactions on}, vol.~15, no.~3, pp. 710--715,
  2013.

\bibitem{sezer2013we}
S.~Sezer, S.~Scott-Hayward, P.-K. Chouhan, B.~Fraser, D.~Lake, J.~Finnegan,
  N.~Viljoen, M.~Miller, and N.~Rao, ``Are we ready for sdn? implementation
  challenges for software-defined networks,'' \emph{Communications Magazine,
  IEEE}, vol.~51, no.~7, 2013.

\bibitem{civanlar2010qos}
S.~Civanlar, M.~Parlakisik, A.~M. Tekalp, B.~Gorkemli, B.~Kaytaz, and E.~Onem,
  ``A {{QoS}} -enabled openflow environment for scalable video streaming,'' in
  \emph{GLOBECOM Workshops (GC Wkshps), 2010 IEEE}.\hskip 1em plus 0.5em minus
  0.4em\relax IEEE, 2010, pp. 351--356.

\bibitem{Armbrust2010}
\BIBentryALTinterwordspacing
M.~Armbrust, A.~Fox, R.~Griffith, A.~D. Joseph, R.~Katz, A.~Konwinski, G.~Lee,
  D.~Patterson, A.~Rabkin, I.~Stoica, and M.~Zaharia, ``A view of cloud
  computing,'' \emph{Commun. ACM}, vol.~53, no.~4, pp. 50--58, Apr. 2010.
  [Online]. Available: \url{http://doi.acm.org/10.1145/1721654.1721672}
\BIBentrySTDinterwordspacing

\bibitem{capwap2009}
P.~Calhoun, ``Rfc 5415. {Control and Provisioning of Wireless Access Points
  (CAPWAP) Protocol Specifications},''
  \emph{\url{https://tools.ietf.org/rfc/rfc5415.txt}}, 2009.

\bibitem{meraki}
``{Meraki Networks [Online]},'' \url{http://meraki.cisco.com}, accessed:
  2014-06-1.

\bibitem{aruba}
``{Aruba Networks [Online]},'' \url{http://cloud.arubanetworks.com/}, accessed:
  2014-06-1.

\bibitem{engelstad2005non}
P.~E. Engelstad and O.~N. Osterbo, ``Non-saturation and saturation analysis of
  {IEEE} 802.11e {EDCA} with starvation prediction,'' in \emph{Proceedings of
  the 8th ACM international symposium on Modeling, analysis and simulation of
  wireless and mobile systems}.\hskip 1em plus 0.5em minus 0.4em\relax ACM,
  2005, pp. 224--233.

\bibitem{malone2007mac}
D.~Malone, P.~Clifford, and D.~J. Leith, ``{MAC} layer channel quality
  measurement in 802.11,'' \emph{IEEE Communications Letters}, vol.~11, no.~2,
  pp. 143--145, 2007.

\bibitem{giustiniano2010measuring}
D.~Giustiniano, D.~Malone, D.~J. Leith, and K.~Papagiannaki, ``Measuring
  transmission opportunities in 802.11 links,'' \emph{IEEE/ACM Transactions on
  Networking (TON)}, vol.~18, no.~5, pp. 1516--1529, 2010.

\bibitem{wu2006ieee}
H.~Wu, A.~Wang, Q.~Zhang, and X.~Shen, ``{IEEE} 802.11e enhanced distributed
  channel access {EDCA} throughput analysis,'' in \emph{Communications, 2006.
  ICC'06. {IEEE} International Conference on}, vol.~1.\hskip 1em plus 0.5em
  minus 0.4em\relax IEEE, 2006, pp. 223--228.

\bibitem{kosek2011simple}
K.~Kosek-Szott, M.~Natkaniec, and A.~R. Pach, ``A simple but accurate
  throughput model for {IEEE} 802.11 {EDCA} in saturation and non-saturation
  conditions,'' \emph{Computer Networks}, vol.~55, no.~3, pp. 622--635, 2011.

\bibitem{akyildiz2006next}
I.~Akyildiz, W.~Lee, M.~Vuran, and S.~Mohanty, ``Next generation/dynamic
  spectrum access/cognitive radio wireless networks: a survey,'' \emph{Computer
  Networks}, vol.~50, no.~13, pp. 2127--2159, 2006.

\bibitem{haykin2005cognitive}
S.~Haykin, ``Cognitive radio: brain-empowered wireless communications,''
  \emph{Selected Areas in Communications, {IEEE} Journal on}, vol.~23, no.~2,
  pp. 201--220, 2005.

\bibitem{clark2003knowledge}
D.~D. Clark, C.~Partridge, J.~C. Ramming, and J.~T. Wroclawski, ``A knowledge
  plane for the internet,'' in \emph{Proceedings of the 2003 conference on
  Applications, technologies, architectures, and protocols for computer
  communications}.\hskip 1em plus 0.5em minus 0.4em\relax {ACM}, 2003, pp.
  3--10.

\bibitem{fette2009cognitive}
B.~A. Fette, \emph{Cognitive radio technology}.\hskip 1em plus 0.5em minus
  0.4em\relax Access Online via Elsevier, 2009.

\bibitem{thomas2007cognitive}
R.~W. Thomas, D.~H. Friend, L.~A. DaSilva, and A.~B. MacKenzie, \emph{Cognitive
  networks}.\hskip 1em plus 0.5em minus 0.4em\relax Springer, 2007.

\bibitem{berlemann2005radio}
L.~Berlemann, G.~R. Hiertz, B.~H. Walke, and S.~Mangold, ``Radio resource
  sharing games: enabling {{QoS}} support in unlicensed bands,'' \emph{Network,
  IEEE}, vol.~19, no.~4, pp. 59--65, 2005.

\bibitem{xing2007dynamic}
Y.~Xing, C.~N. Mathur, M.~A. Haleem, R.~Chandramouli, and K.~Subbalakshmi,
  ``Dynamic spectrum access with {{QoS}} and interference temperature
  constraints,'' \emph{Mobile Computing, {IEEE} Transactions on}, vol.~6,
  no.~4, pp. 423--433, 2007.

\bibitem{su2008cross}
H.~Su and X.~Zhang, ``Cross-layer based opportunistic {MAC} protocols for
  {{QoS}} provisionings over cognitive radio wireless networks,''
  \emph{Selected Areas in Communications, {IEEE} Journal on}, vol.~26, no.~1,
  pp. 118--129, 2008.

\bibitem{berlemann2006coexistence}
L.~Berlemann, C.~Hoymann, G.~R. Hiertz, and S.~Mangold, ``Coexistence and
  interworking of {IEEE} 802.16 and {IEEE} 802.11(e),'' in \emph{Vehicular
  Technology Conference, 2006. VTC 2006-Spring. {IEEE} 63rd}, vol.~1.\hskip 1em
  plus 0.5em minus 0.4em\relax IEEE, 2006, pp. 27--31.

\bibitem{andrews2014will}
J.~G. Andrews, S.~Buzzi, W.~Choi, S.~Hanly, A.~Lozano, A.~C. Soong, and J.~C.
  Zhang, ``What will {5G} be?'' \emph{To appear in {IEEE} JSAC, arXiv preprint
  arXiv:1405.2957}, 2014.

\bibitem{wen2013wirelessbook}
H.~Wen, P.~K. Tiwary, and T.~Le-Ngoc, \emph{Wireless Virtualization}.\hskip 1em
  plus 0.5em minus 0.4em\relax SpringerBriefs in Computer Science, Springer,
  2013.

\bibitem{qadir2013artificial}
J.~Qadir, ``Artificial intelligence based cognitive routing for cognitive radio
  networks,'' \emph{arXiv preprint arXiv:1309.0085}, 2013.

\bibitem{how2011routing}
K.~C. How, M.~Ma, and Y.~Qin, ``Routing and qos provisioning in cognitive radio
  networks,'' \emph{Computer Networks}, vol.~55, no.~1, pp. 330--342, 2011.

\bibitem{chowdhury2009tp}
K.~R. Chowdhury, M.~Di~Felice, and I.~F. Akyildiz, ``Tp-crahn: A transport
  protocol for cognitive radio ad-hoc networks,'' in \emph{INFOCOM 2009,
  IEEE}.\hskip 1em plus 0.5em minus 0.4em\relax IEEE, 2009, pp. 2482--2490.

\end{thebibliography}

%
%
%
%

\end{document}